\documentclass[prd,aps,twocolumn,a4paper,showkeys,nofootinbib]{revtex4-1}

\usepackage{graphicx,psfrag}
\usepackage{mathrsfs}
\usepackage{amsmath,amsfonts,amssymb}
\usepackage{multirow}
\usepackage{comment}
\usepackage{ulem}
\usepackage{multirow}
\usepackage{hyperref}
\usepackage{acronym}
\usepackage{longtable}

\newcommand{\be}{\begin{equation}}
\newcommand{\ee}{\end{equation}}
\newcommand{\bea}{\begin{eqnarray}}
\newcommand{\eea}{\end{eqnarray}}
\newcommand{\bel}{\begin{align}}
\newcommand{\eel}{\end{align}}

\def\data{{\text{\textbf d}}}

\def\half{\frac{1}{2}}
\def\e{{\rm e}}
\def\i{{\rm i}}
\def\d{{\rm d}}

\def\Msun{{\rm M_{\odot}}}
\def\Mmax{{M_{\rm max}^{\rm TOV}}}

\def\Rmax{{R_{\rm max}^{\rm TOV}}}
\def\rhomax{{\rho_{\rm max}^{\rm TOV}}}

\def\Fbar{{\bar{\mathcal{F}}}}

\def\freeparams{{\params_{\rm PM}^{\rm free}}}
\def\extparams{{\params_{\rm ext}}}
\def\Mc{\mathcal{M}}
\def\kt{\kappa^{\rm T}_2}

\def\params{\boldsymbol{\theta}}
\def\recalib{\boldsymbol{\delta}}
\def\GMc2{{\rm G M_{\odot} c^{-2}}}

\def\wavelet{{W}}

\def\core{{\scshape CoRe}}

\def\model{{\tt NRPMw}}

\newcommand\teob[1]{{\tt TEOBResumS{#1}}}
\def\teobspa{\teob{-SPA}}
\def\teobspa{{\tt TEOBResumSPA}}

\def\teobnrpmw{{\teobspa{\tt\_}\model}}

\def\bajes{{\tt bajes}}

\def\spin{{\boldsymbol \chi}}
\def\chieff{{\chi_{\rm eff}}}
\def\adrift{\alpha_{\rm drift}}
\def\ie{\textit{i.e.}}
\def\eg{\textit{e.g.}}

\usepackage{pifont} % \cmark \xmark

\usepackage{color}
\definecolor{cyan}{rgb}{0,0.9,0.9}
\definecolor{orange}{rgb}{0.9,0.5,0}
\definecolor{magenta}{rgb}{1,0,1}
\definecolor{purple}{rgb}{0.8,0.4,0.8}
\definecolor{gray}{rgb}{0.8242,0.8242,0.8242}

\newcommand{\paperI}{paper~{I}}
\newcommand{\paperII}{paper~{II}}

\newacro{adm}[ADM]{Arnowitt-Deser-Misner}
\newacro{bbh}[BBH]{binary black hole}
\newacro{bh}[BH]{black hole}
\newacroplural{bh}[BHs]{black holes}
\newacro{bhns}[BHNS]{black hole-neutron star}
\newacro{bns}[BNS]{Binary neutron star}
\newacro{bf}[BF]{Bayes' factor}
\newacro{cbc}[CBC]{compact binary coalescence}
\newacro{ce}[CE]{Cosmic Explorer}
\newacro{cl}[CL]{confidence level}
\newacro{da}[DA]{data analysis}
\newacro{et}[ET]{Einstein Telescope}
\newacro{emri}[EMRI]{extreme mass ratio inspiral}
\newacro{eob}[EOB]{effective-one-body}
\newacro{eos}[EoS]{equation of state}
\newacro{eom}[EOM]{equations of motion}
\newacro{fd}[FD]{frequency domain}
\newacro{fft}[FFT]{Fast Fourier transform}
\newacro{gw}[GW]{{gravitational wave}}
\newacroplural{gw}[GWs]{Gravitational waves}
\newacro{gr}[GR]{general relativity}
\newacro{grb}[GRB]{gamma-ray burst}
\newacro{grhd}[GRHD]{general-relativistic hydrodynamics}
\newacro{gwosc}[GWOSC]{Gravitational Wave Open Science Center}
\newacro{gwtc1}[GWTC-1]{the first gravitational-wave transients catalog}
\newacro{gsf}[GSF]{Gravitational Self Force}
\newacro{hm}[HM]{Higher mode}
\newacroplural{hm}[HMs]{Higher modes}
\newacro{ifo}[IFO]{interferometer}
\newacro{imr}[IMR]{inspiral-merger-ringdown}
\newacro{im}[IM]{inspiral merger}
\newacro{impm}[IMPM]{inspiral-merger-postmerger}
\newacro{kagra}[KAGRA]{Kamioka Gravitational Wave Detector}
\newacro{ligo}[LIGO]{Laser Interferometer Gravitational-Wave Observatory}
\newacro{lso}[LSO]{Last Stable Orbit}
\newacro{lvc}[LVC]{LIGO-Virgo Collaboration}
\newacro{lvk}[LVK]{LIGO-Virgo-Kagra Collaboration}
\newacro{lo}[LO]{leading order}
\newacro{ns}[NS]{neutron star}
\newacroplural{ns}[NSs]{neutron stars}
\newacro{nr}[NR]{numerical relativity}
\newacro{nqc}[NQCs]{next-to-quasicircular corrections}
\newacro{nlo}[NLO]{next-to-leading order}
\newacro{nnlo}[NNLO]{next-to-next-to-leading order}
\newacro{n3lo}[N3LO]{next-to-next-to-next-to-leading order}
\newacro{n4lo}[N3LO]{next-to-next-to-next-to-next-to-leading order}
\newacro{ode}[ODE]{Ordinary Differential Equation}
\newacroplural{ode}[ODEs]{Ordinary Differential Equations}
\newacro{pc}[PC]{prompt collapse}
\newacro{pe}[PE]{parameter estimation}
\newacro{pn}[PN]{post-Newtonian}
\newacro{pm}[PM]{postmerger}
\newacro{psd}[PSD]{power spectral density}
\newacro{pa}[PA]{post-adiabatic}
\newacro{qnm}[QNM]{quasi-normal mode}
\newacro{qc}[QC]{quasicircular}
\newacro{rwz}[RWZ]{Regge-Wheeler-Zerilli}
\newacro{sm}[SM]{Supplemental Material}
\newacro{snr}[SNR]{signal-to-noise ratio}
\newacro{spa}[SPA]{stationary-phase approximation}
\newacro{sxs}[SXS]{Simulating eXtreme Spacetimes}
\newacro{td}[TD]{time domain}
\newacro{tov}[TOV]{Tolmann-Oppenheimer-Volkoff}
\newacro{xg}[XG]{next-generation}

\begin{document}

\title{Kilohertz Gravitational Waves from Binary Neutron Star Mergers:\\
Full Spectrum Analyses and High-density Constraints on Neutron Star Matter} 

\author{Giulia \surname{Huez}$^{1}$}
\author{Sebastiano \surname{Bernuzzi}$^{1}$}
\author{Matteo \surname{Breschi}$^{1}$}
\author{Rossella \surname{Gamba}$^{2,3}$}
\affiliation{${}^1$Theoretisch-Physikalisches Institut, Friedrich-Schiller-Universit{\"a}t Jena, 07743, Jena, Germany\\
				${}^2$Institute for Gravitation and the Cosmos, The Pennsylvania State University, University Park, Pennsylvania 16802, USA \\
				${}^3$Department of Physics, University of California, Berkeley, California 94720, USA}

\date{\today}

\begin{abstract}
  We demonstrate Bayesian analyses of the complete gravitational-wave spectrum of binary neutron
  star mergers events with the next-generation detector Einstein Telescope.
  Our mock analyses are performed for 20 different signals using the {\teobnrpmw} waveform
  that models gravitational waves from the inspiral to the postmerger phase.
  They are employed to validate a pipeline for neutron star's extreme
  matter constraints with prospective detections and under minimal
  hypotheses on the equation of state.
  The proposed analysis stack 
  delivers inferences for the mass-radius curve,
  the mass dependence of the quadrupolar tidal polarizability parameter, the neutron
  star's maximum density, the maximum mass and the relative radius, and the
  pressure-density relation itself. 
  We show that a single event at a signal-to-noise ratio close to the minimum 
  threshold for postmerger detection is sufficient to tightly
  constrain all the above relations as well as quantities like the maximum
  mass (maximum density) to precision of ${\sim}6$\%
  (${\sim}10$\%) at 90\% credibility level.
  We also revisit inferences of prompt black hole formation with full spectrum signals 
  and find that the latter can be robustly identified, even 
  when the postmerger is not detectable due to a low signal-to-noise ratio.
  New results on the impact of the initial signal frequency and of the 
  detector configuration (triangular vs. two-L) on the  
  source's parameters estimation are also reported.   An improvement of approximately one order of magnitude in
  the precision of the chirp mass and mass ratio can be achieved by lowering the initial 
  frequency from 20~Hz to 2~Hz. The two-L configuration shows instead significant improvements 
  on the inference of the source declination, due to geographical separation of the two detectors.
\end{abstract}


\maketitle

\section{Introduction}
\label{sec:intro}

\acp{bns} are the main sources for \ac{xg} ground-based \ac{gw} detectors, such as \ac{et} 
\cite{Hild:2010id,Hild:2011np,Abac:2025saz,Maggiore:2019uih,Punturo:2010zza} 
and \ac{ce}~\cite{Reitze:2019iox,Evans:2021gyd}. They are expected to be observed with a rate of 
$7 \times 10^4$ mergers per year~\cite{Belgacem:2019tbw,Kalogera:2021bya,Iacovelli:2022bbs}.
\ac{et} is designed with a xylophone configuration, combining a cryogenic interferometer 
for the low-frequency band with a high-power interferometer targeting frequencies above 
${\sim}100~{\rm Hz}$. This design will not only significantly enhance current \ac{gw} 
detector sensitivity, but also enable the observations
of \ac{pm} signals at kiloHertz frequencies.
\acp{gw} observations of the full spectrum (inspiral-merger-postmerger) of a \ac{bns} merger 
can convey unique information on the nuclear matter that constitutes this compact 
object~\eg~\citep{Bauswein:2017vtn,Breschi:2019srl,Radice:2020ddv,Breschi:2021xrx,Wijngaarden:2022sah},
thus improving current GW constraints after GW170817~\cite{LIGOScientific:2017vwq,LIGOScientific:2017ync,LIGOScientific:2018cki}.
While the \ac{im} carries the footprint of the matter 
properties at the progenitor densities, \ie~$\rho\lesssim {2}\,\rho_{\rm sat}$ 
(where $\rho_{\rm sat}\simeq 2.7{\times }10^{14}~{\rm g}~{\rm cm}^{-3}$
is the nuclear saturation density), the 
kiloHertz \ac{pm} transient encodes information about the high-density
(${\sim}3-6\rho_{\rm sat}$) regime of the \ac{eos}~\cite{Breschi:2021xrx}, 
enabling the investigation of unrevealed properties of the nuclear matter 
at high densities, unreachable with modern experiments~\cite{Radice:2016rys,Bauswein:2018bma,Breschi:2021xrx,Most:2021ktk,Prakash:2021wpz}.

Inference of source properties and constraints on the \ac{ns} \ac{eos} requires the availability 
of complete waveform templates for matched-filtering analyses. 
These templates are crucially designed using empirical relations computed from
\ac{nr} simulations and connecting morphological features of the \ac{pm} signal 
to \ac{ns} properties, \eg~\cite{Shibata:2002jb,Stergioulas:2011gd,Bauswein:2011tp,Bauswein:2012ya,Hotokezaka:2013iia,Takami:2014zpa,Bauswein:2015yca,Bernuzzi:2015rla,Breschi:2021xrx,Breschi:2022xnc}. 
In particular, \citet{Breschi:2019srl} proposed the first time-domain full spectrum model that 
was constructed by combining an \ac{eob} \ac{im} waveform with a
\ac{nr}-informed analytical \ac{pm} model (see also~\cite{Breschi:2022xnc,Breschi:2022ens}). Similar 
approaches have been used in~\cite{Wijngaarden:2022sah,Puecher:2022oiz}, where 
phenomenological frequency-domain \ac{im} waveforms are stitched to \ac{pm} models based on 
wavelets or Lorenzian templates. In all cases the phenomenological \ac{pm} parameters are 
connected to the source parameter using \ac{nr} empirical relations; see also 
\citep{Bauswein:2015vxa,Bose:2017jvk,Soultanis:2021oia,Vretinaris:2025wdu} for other examples 
of analytical \ac{pm} models and~\citep{Easter:2020ifj,Whittaker:2022pkd} for data-driven 
\ac{pm} models, which have not been incorporated in full spectrum models yet. 

These modeling efforts, combined with the exceptional sensitivity of \ac{xg} detectors,
provide a crucial foundation for accurate \ac{eos} inference. 
Notably, even a single detection of \ac{bns} with \ac{et} may 
suffice to place stringent constraints on the \ac{ns} mass-radius relation~\cite{Breschi:2021xrx}.
That approach enabled tight constraints of the mass-radius diagram and demonstrated that the 
maximum \ac{ns} mass can be determined with an accuracy better than 12\%.
Additional works~\cite{Breschi:2023mdj,Rezzolla:2016nxn,Chatziioannou:2017ixj,Lehoucq:2025ruc,Vretinaris:2025wdu}
have similarly combined information from the \ac{im} section to \ac{pm} features, highlighting 
the complementarity of these two regimes in probing supranuclear matter.
Numerous proof of principle studies focus exclusively on the \ac{pm} \ac{gw} emission 
from \acp{bns} to investigate the high-density nuclear matter through \ac{nr} simulations. 
These works exploit the connection between peak frequencies and \ac{eos}-dependent 
properties, such as total mass and compactness~\cite{Bauswein:2014qla,Most:2021ktk,Easter:2018pqy,Takami:2014tva,Bauswein:2015vxa,Ecker:2024uqv,Bamber:2024qzi,Mitra:2025uka}.
Furthermore, recent works explored the potential for \ac{eos} inference
using advanced and \ac{xg} detectors~\cite{Walker:2024loo,Gupta:2022qgg,Iacovelli:2023nbv,Bandopadhyay:2024zrr,Kiendrebeogo:2023hzf}
and assess the impact of different waveform models on nuclear \ac{eos} inference~\cite{Yelikar:2024rmh}.

In the cases where the presence of the \ac{pm} is not detectable, a major issue is the inference 
of prompt (rapid) \ac{bh} formation after merger. While electromagnetic counterparts can provide 
valuable information in this regard~\cite{LIGOScientific:2017ync,Coulter:2017wya,Chornock:2017sdf,Nicholl:2017ahq,Cowperthwaite:2017dyu,Tanvir:2017pws,Tanaka:2017qxj}, 
\acp{gw} remain the key messenger to identify unambiguously the presence of a remnant \ac{bh}.
Many authors have explored the conditions under which a \ac{pc} occurs instead of the formation of a \ac{ns} remnant. 
This is primarily connected to the total mass of the \ac{bns} and the \ac{eos} of 
dense nuclear matter~\cite{Shibata:2005ss,Hotokezaka:2011dh,Bauswein:2013jpa,Kashyap:2021wzs,Perego:2021mkd}.
In addition, \ac{pc} analysis can be employed to place constraints 
on the \ac{eos}~\cite{Bauswein:2017vtn,Bauswein:2020aag,Tootle:2021umi,Kashyap:2021wzs}.
\citet{Agathos:2019sah} presented the first comprehensive Bayesian analysis to assess the 
probability of \ac{pc} directly from the inspiral \ac{gw} signal. This methodology was subsequently
applied in the analysis of GW190425~\cite{LIGOScientific:2020aai}.

In this paper, we perform an extensive investigation of the capabilities of 
\ac{et} to infer \ac{bns} properties. We firstly focus on the low-frequency regime, assessing 
the impact on source parameter estimation of the initial frequency and of  
detector configurations (two-L and triangular).
We then concentrate in the 
detections of the complete \ac{gw} spectrum and study the prospective accuracy of \ac{eos} inference.
We develop and employ a full spectrum model which combines the fast frequency-domain \ac{eob} 
{\teobspa}~\cite{Akcay:2018yyh, Gamba:2020ljo, Gamba:2023mww} with the frequency-domain wavelet-based {\model} 
model~\cite{Breschi:2022xnc,Breschi:2022ens} ({\paperI} and {\paperII} hereafter). 
We then perform Bayesian mock analyses of 20 fiducial \acp{bns}, covering the frequency range 
from $f_0=20$~Hz up to the kiloHertz regime. These analyses serve not only to demonstrate 
the feasibility of combining information across the full \ac{gw} spectrum, from inspiral 
to merger and postmerger, but also to quantify the achievable accuracy in the 
inference of source parameters. Beyond parameter estimation, our primary objective is to 
assess the improvements in constraints on the supranuclear properties governing \ac{ns} 
matter. By incorporating both low- and high-frequency components of the signal, we 
explore how analyses of \ac{et} signals enhance our ability to probe the dense-matter regime, 
including tighter bounds on the mass-radius relation, the maximum mass supported by a 
stable nonrotating \ac{ns} and the \ac{pc} inference.

The paper is structured as follows. In Sec.~\ref{sec:method} we introduce the full spectrum 
waveform model and the framework used for the analyses. We present the results in 
Sec.~\ref{sec:results}, covering both inspiral only and full spectrum analyses, as well as 
a discussion about the \ac{eos} constraints and the probability of prompt \ac{bh} formation. 
Finally Sec.~\ref{sec:conclusion} provides a summary and our main conclusion. 
Moreover, we include Appendices about the waveform model and Bayesian analysis.

\paragraph*{Conventions --}
All quantities are expressed in SI units, with masses in solar masses $\Msun$ and distances 
in Mpc. The symbol $\log$ denotes the natural logarithm.
The waveform strain is decomposed in multipoles $(\ell,m)$ as 
\be
\label{eq:hdecomp}
h_+ - \i h_\times
=D_L^{-1}\sum_{\ell=2}^\infty\sum_{m=-\ell}^{\ell} h_{\ell m}(t)\,{}_{-2}Y_{\ell m}(\iota,\varphi),
\ee
where $D_L$ is the luminosity distance, ${}_{-2}Y_{\ell m}$ are the $s=-2$ spin-weighted spherical harmonics 
and $\iota,\varphi$ are, respectively, the polar and azimuthal angles that define the orientation of the binary 
with respect to the observer. 
Each waveform mode $h_{\ell m}(t)$ is decomposed 
in amplitude $A_{\ell m}(t)$ and phase $\phi_{\ell m}(t)$, as
\be
\label{eq:hlm}
h_{\ell m}(t) = A_{\ell m}(t)e^{- \i \phi_{\ell m}(t)} \,,
\ee
with a related \ac{gw} frequency,
\be
\label{eq:fgw}
\omega_{\ell m}(t) =2\pi f_{\ell m}(t) = \frac{\d}{\d t}{\phi_{\ell m}}(t) \, .
\ee
The moment of merger is defined as the time of the peak of $A_{22}(t)$. 
When the $(\ell, m)$ indices are omitted, we consider only the dominant $(2,2)$ mode. 

The waveform depends on intrinsic and extrinsic source parameters. The intrinsic parameters of 
a \ac{bns} system are masses, spins and tidal polarizability parameters, 
\eg~$\params_{\rm bin}=\{M,q,\chi_{1},\chi_{2},\Lambda_1,\Lambda_2\}$. The total binary mass 
is indicated with $M= m_1 + m_2$, the mass ratio is $q = m_1 /m_2 \ge 1$, and the symmetric mass 
ratio is $\nu = m_1 m_2 / M^2$. The dimensionless spin vectors are denoted with $\spin_i$ 
for $i=1,2$ and the spin component aligned with the orbital angular momentum $\textbf{L}$ 
are labeled as $\chi_i = \spin_{i}\cdot \textbf{L} / |\textbf{L}|$.
The effective spin parameter $\chieff$ is a mass-weighted, aligned spin combination, i.e.,
\be
\label{eq:chieff}
\chieff = \frac{m_1 \chi_{1}+m_2 \chi_{2}}{M}\,.
\ee
The quadrupolar tidal polarizability parameters are defined as 
$\Lambda_{i}=({2}/{3})\,k_{2,i}\,C_i^{-5}$ for $i=1,2$, where $k_{2,i}$ 
and $C_i$ are the second Love number and the compactness of the $i$th object, 
respectively. 
The reduced tidal parameter $\tilde{\Lambda}$ is
\be
\label{eq:lamtilde}
\tilde{\Lambda} = \frac{16}{13}\left[ \frac{(m_1+12m_2)m_1^4\Lambda_1}{M^5} + (1 \leftrightarrow 2)\right]\,.
\ee
The quadrupolar tidal polarizability $\kt$ is
\be
\label{eq:k2t}
\kt =3\nu\,\left[\left(\frac{m_1}{M}\right)^3 \Lambda_1 + (1\leftrightarrow 2)\right]\,.
\ee
The extrinsic parameters of the source 
$\extparams=\{D_L, \iota , \alpha, \delta, \psi, t_{\rm mrg},\phi_{\rm mrg}\}$, \ie~inclination 
angle $\iota$, right ascension angle $\alpha$, declination angle $\delta$, polarization angle 
$\psi$, time of coalescence $t_{\rm mrg}$, and phase at the merger $\phi_{\rm mrg}$, allow us 
to identify the location and orientation of the source.
Moreover, we include the recalibration parameters $\recalib_{\rm fit}$ that account for deviations 
from the predictions of the \ac{eos}-insensitive relations consistently with the 
related theoretical uncertainties. 
{\model} is parametrized by two additional sets of degrees of freedom (see {\paperI} for a 
detailed discussion). The \ac{pm} parameters $\freeparams=\{\phi_{\rm PM},t_c,\adrift\}$ 
correspond to: the \ac{pm} phase shift $\phi_{\rm PM}$ that identifies the phase discontinuity 
after merger; the time of collapse $t_c$ that characterizes the collapse of the remnant into 
\ac{bh}; and the frequency drift $\adrift$ that accounts for linear evolution of the 
dominant $f_2$ component.

\section{Method}
\label{sec:method}

\subsection{Waveform model}
\label{sec:method:wvf}

\begin{figure}[t]
  \centering 
  \includegraphics[width=0.49\textwidth]{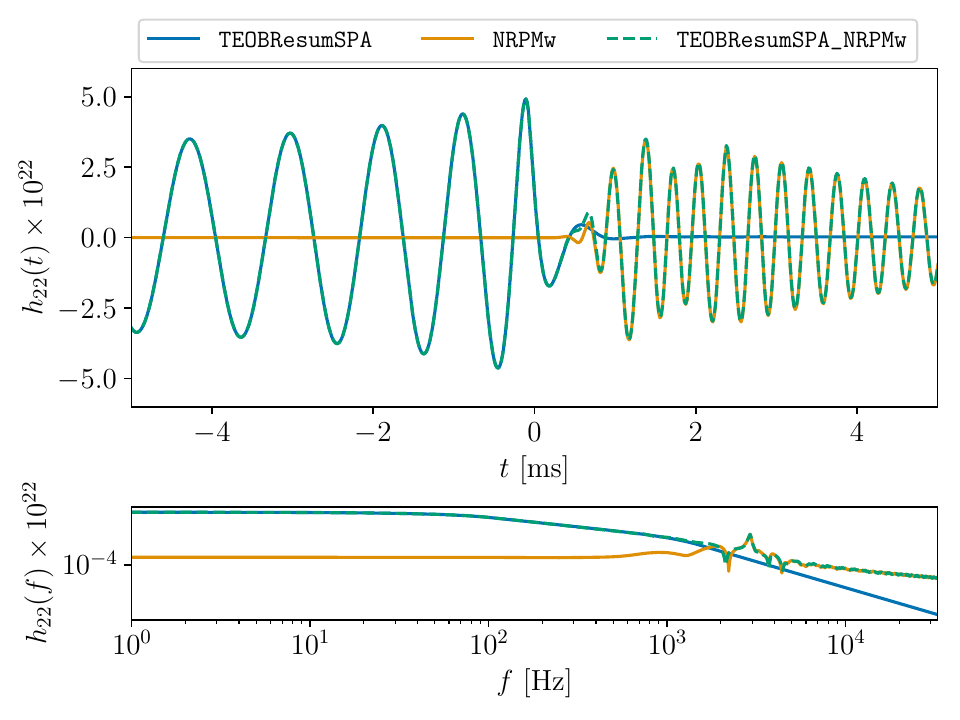}
    \caption{Attachment between {\teobspa} and \model~for a \ac{gw} template 
  of nonspinning binary with
    $m_1=1.5~\Msun$, 
    $m_2=1.4~\Msun$, 
    $\Lambda_1=400$, 
    $\Lambda_2=600$
    and $D_L=40~{\rm Mpc}$.
	Since the model is defined in the frequency domain, we obtain $h_{22}(t)$ by performing
	an inverse \ac{fft} of $h_{22}(f)$.
  }
  \label{fig:attach}
\end{figure}

In paper I we introduced the \ac{pm} \model~waveform, which can be employed to
extend in the kiloHertz regime the existing frequency-domain approximants for \ac{gw} signals 
from \ac{bns} mergers. For our studies, we resort to the \ac{eob} model 
\teob{}~\cite{Nagar:2018zoe, Akcay:2018yyh, Nagar:2018plt}, in its frequency domain formulation, 
{\teobspa}~\cite{Nagar:2018gnk,Gamba:2020ljo}, for the representation of the \ac{im} part and
present here the full spectrum template {\teobnrpmw}.

We attach the \ac{eob} model for the \ac{im} to the \ac{pm} \model~template, focusing on the dominant $(2,2)$ mode. 
The information required to perform the attachment of {\model} with {\teobspa} is the time and the phase at merger, 
respectively $t_{\rm mrg}$ and $\phi_{\rm mrg}$. These data can be extracted from the \ac{eob} \ac{im} waveform. Then, 
resorting linear properties of the Fourier transform, the \ac{impm} waveform can be computed as the sum of the premerger 
template $h_{\rm EOB}$ and the \ac{pm} one $h_{\rm PM}$ in the frequency-domain, 
\be
\label{eq:imp}
h_{\rm IMPM}(f) = h_{\rm EOB}(f) + h_{\rm PM}(f)\,\e^{-2\pi\i f t_{\rm mrg} + \phi_{\rm mrg}}\,,
\ee
where $h_{\rm PM}(f)$ is taken such that its reference time and phase are vanishing.

Figure~\ref{fig:attach} illustrates an example of such an attachment procedure for a representative 
configuration with $M=2.9 \Msun$, $q=1.072$, $\Lambda_1=400$ and $\Lambda_2=600$.
As shown, the {\teobnrpmw}~waveform (dashed line) accurately reproduces the \ac{im} dynamics as 
described by the {\teobspa}~template, as well as the \ac{pm} segment following \model.
The attachment occurs after the merger, with the adoption of the phase from the \ac{eob} model, 
according to Eq.~\eqref{eq:imp}.
To guarantee a smooth transition, the fusion wavelet, \ie~the portion of the \model~waveform between the merger time 
(as predicted by {\teobspa}) and $t_0$, defined as the first amplitude's minimum after the 
merger (See paper I), is used to perform a continuous attachment of the \ac{pm} with the 
\ac{im}. Consequently, this segment is not compatible with either model.

We also validate the model against the $(2,2)$-mode of six \ac{nr} simulations that are not included in the 
calibration set of \model, finding that the model is $89-95\%$ faithful to \ac{nr} when 
considering the full spectrum of frequencies and the ET-D sensitivity curve~\cite{Hild:2010id,Hild:2011np}. 
More details on the \ac{impm} model and its validation can be found in Appendix~\ref{app:wvf}.

\subsection{Inference}
\label{sec:method:inf}

\begin{figure}[t]
  \centering 
  \includegraphics[width=0.49\textwidth]{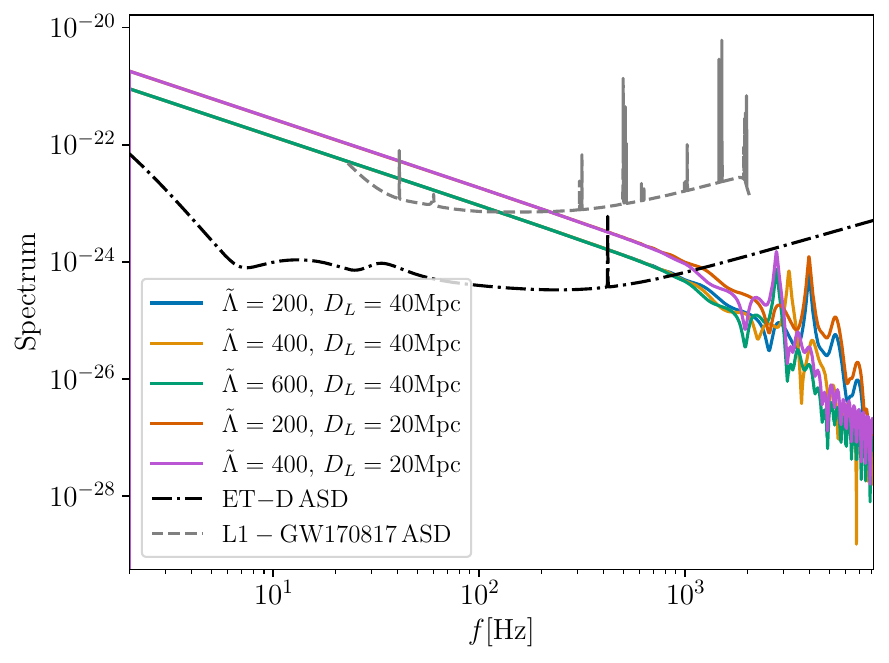}
  \caption{Amplitude spectral density (ASD) of \ac{et} in configuration D~\cite{Hild:2010id,Hild:2011np} (in black) and 
	comparison with ASD of LIGO L1 detector, relative to GW170817 event~\cite{LIGOScientific:2018mvr} (in gray), 
	to show the improvement in the detection of \ac{gw} signals with \ac{xg} detectors.
	The solid colored lines represent five exemplary \ac{bns} waveform spectra with {\teobnrpmw} template for
    $\Mc=1.24~\Msun$, $q=1$, varying $\tilde{\Lambda}$ and $D_L$ as indicated in the legend.}
  \label{fig:asd}
\end{figure}

We perform Bayesian \ac{pe} experiments by injecting and analyzing mock zero noise \ac{bns} 
signals with \ac{xg} detectors. We employ the \ac{psd} curve of the \ac{et} detector, as reported in 
Fig.~\ref{fig:asd}, in configuration D~\cite{Hild:2010id,Hild:2011np}, \ie~an 
equilateral triangular shape with $10~{\rm km}$ long arms. We simulate both 
\ac{im}-only and \ac{impm} waveforms and recover them using the same model employed 
for the injection, {\teobspa} and {\teobnrpmw} respectively. 

Bayesian inference is performed with the parallel \bajes~pipeline~\cite{Breschi:2021wzr}. 
In order to efficiently explore the parameter space and compute the evidence, we utilize 
nested sampling algorithm~\cite{Veitch:2014wba,Handley:2015fda,Allison:2013npa} as implemented 
in the {\tt dynesty} library~\cite{Speagle:2019ivv} with a minimum of 2048 live points.

Bayesian analysis is based on the evaluation of the likelihood function (See Appendix~\ref{app:inf:bajes}
for additional details), which is the most computationally expensive part. 
To speed up the computations, we make use of the relative binning technique~\cite{Dai:2018dca, Zackay:2018qdy} 
in the \ac{im} part. Relative binning consists in the precomputation of frequency-binned overlap of a fiducial 
waveform with the data, since the ratio of neighboring waveforms in frequency domain is smooth in the inspiral part. 
In particular, for the \ac{im}-only \acp{pe} we employ the relative binning method for the whole frequency range of the integration. 
For the \ac{impm} cases, we utilize relative binning only up to a cutoff frequency $f_{\rm cut}=2048~{\rm Hz}$ 
and we perform a normal Bayesian \ac{pe} without speed up after the cutoff frequency.
The two likelihoods are then combined by summing their respective contributions
\be
	\log p(\data|\params,H_{\rm S}) = \log p(\data_{\rm IM}|\params,H_{\rm IM}^{\rm RB}) + \log p(\data_{\rm PM}|\params,H_{\rm PM}).
\label{eq:likel_sum}
\ee
where $\log p(\data_{\rm IM}|\params,H_{\rm IM}^{\rm RB})$ is the likelihood evaluated up to $f_{\rm cut}$ with relative binning, 
while $\log p(\data_{\rm PM}|\params,H_{\rm PM})$ is evaluated in the high-frequency section, 
both using {\teobnrpmw} model.
In order to investigate how sensitive the choice of $f_{\rm cut}$ is, we run 3 additional \ac{impm} analyses 
with different \acp{eos} with $f_{\rm cut} = 1024~{\rm Hz}$ and compare posteriors with the ones of the analyses 
presented in Sec.~\ref{sec:results}. In all the cases the posteriors are completely compatible, we
therefore conclude that the cutoff frequency used does not have any impacts in the result of the inference.
For further explanation of the relative binning implementation and a discussion of the 
approximations used, we refer to Appendix~\ref{app:inf:rb}.

\subsection{EoS constraints from quasiuniversal relations}
\label{sec:method:nrrel}
 
Given an estimation of the intrinsic \ac{bns} parameters, we show how 
\ac{xg} detectors can enable precise constraints on the \ac{eos} of \acp{ns}.
Inference of \ac{im} delivers posteriors of the masses of the compact objects and of the 
tidal polarizability parameters.
The information contained in the \ac{impm} signals, instead, allows to go beyond $m_i$ and $\Lambda_i$
and probe high-density \ac{ns} properties. 
\citet{Breschi:2021xrx} combined the Fisher matrix approach in the \ac{im} part with the 
Bayesian analysis of the \ac{pm} in order to obtain a tighter constraint on the mass-radius 
relation. Here we perform a full spectrum Bayesian \ac{pe}, aiming to obtain more reliable
inference, particularly in the high-density regime.
The resulting pipeline constrains different quantities of interest in \ac{eos} modeling and 
is described in what follows.

The first step is to map the posteriors for the total mass and the peak PM frequency $f_2$ 
into posteriors of maximum density $\rhomax$ and radius at the maximum mass 
$\Rmax$, using the \ac{nr} phenomenological relations~\cite{Breschi:2021xrx,Breschi:2024qlc}
\be
\rhomax =  \frac{a_0c^6}{G^3M^2}\left[ 1+a_1\left( \frac{c^3}{\pi G M f_2}\right)^{1/6} \right]\,,
\label{eq:rhomax}
\ee
with $(a_0,a_1)=(0.135905,-0.59506)$ and 
\be
\begin{split}
\Rmax/M &= (5.81\pm 0.13) - (123.4\pm 7.2)Mf_2 \\
&+ (1121\pm 99)(Mf_2)^2\,.
\end{split}
\label{eq:Rmax}
\ee
Subsequently, we use a sample of two millions \acp{eos} from \citet{Godzieba:2020tjn} 
generated under minimal assumptions, \ie~general relativity, causality, the maximum 
\ac{ns} mass above the maximum \ac{ns} mass from pulsar observations, 
$\Mmax > 1.97\Msun$~\cite{Antoniadis:2013pzd}, and consistent with the 
upper bound on the reduced tidal parameter $\tilde{\Lambda}$ from GW170817~\cite{LIGOScientific:2017vwq, LIGOScientific:2018hze, LIGOScientific:2018cki}.
We reweigh the \ac{eos} sample with the posteriors of the masses, the reduced tidal parameter, 
$\rhomax$ and $\Rmax$, employing \texttt{emcee} sampler~\cite{Foreman-Mackey:2012any}. 
This allows us to place constraints not only on the \ac{ns} mass-radius diagram, but also to determine the 
90\% credibility region for the tidal deformability curve $\Lambda(M)$ and on the pressure-density relation $p(\rho)$. 
Specifically, from the \ac{eos} sampling we obtain a distribution of \acp{eos}
that most closely resemble the one used for the injection. For each sampled 
\ac{eos}, we compute the relations between the mass and the radius of the star and 
between the tidal polarizability parameter and the mass, yielding a 90\% \ac{cl} on the 
$M(R)$ and the $\Lambda(M)$ curve, respectively. Using 
the same \ac{eos} distribution we solve the \ac{tov} equations to determine the 
corresponding pressure-density relation, from which we derive the 90\% confidence 
region for the $p(\rho)$ curve.

The maximum mass of nonrotating \acp{ns}, $\Mmax$, is a crucial parameter for 
constraining the mass-radius diagram and consequently the \ac{eos}. Additionally, 
it is also essential for the determination of the probability of prompt collapse.
Specifically, \ac{pc} is expected to occur when the total mass of the system exceeds a 
threshold mass, which is related through phenomenological relations to $\Mmax$~\cite{Hotokezaka:2011dh,Bauswein:2013jpa}
\be
	M_{\rm thr} = k_{\rm thr}(C_{\rm max}) \Mmax,
	\label{eq:Mthr}
\ee
where $k_{\rm thr}(C_{\rm max})$ is a function of the compactness of the star.
The values of $C_{\rm max}$ and $\Mmax$ come from the \ac{eos} sampling 
as presented above. In our analysis we adopt the updated fits for $k_{\rm thr}$ of~\cite{Kashyap:2021wzs},
which were obtained for equal-mass binaries. \citet{Perego:2021mkd} provide a more general 
characterization with mass ratio effects and show that the latter give corrections 
compatible with the fit uncertainty down to mass ratios $q\lesssim1.4-1.5$ 
(depending on \ac{eos}). For larger mass ratio \ac{pc} is driven by accretion effects of 
the secondary stars to the primary and require a more sophisticated parametrization. 
Empirical models for the threshold mass can be obtained also in those cases by considering, 
for example, the \ac{eos} incompressibility parameter at extreme densities. 
Typical deviations from the equal-mass relation are up to ${\sim}20\%$ for mass 
ratios $q\sim2$.

In our \ac{pc} analyses, we consider equal-mass binaries and determine the $90\%$ \ac{cl} 
of the maximum mass $\Mmax$ and the 
corresponding radius $\Rmax$ of the \ac{ns} from the mass-radius diagram and insert
them in Eq.~\eqref{eq:Mthr}. We then calculate the probability of \ac{pc} as
\begin{equation}
	P_{\rm PC} = P(M > M_{\rm thr}|d)\,,
	\label{eq:prob_pc}
\end{equation}
as proposed by \citet{Agathos:2019sah}.

In addition, the presence of \ac{pc} can also be statistically evaluated by computing the \ac{bf} between models that
assume either a \ac{pc} to a \ac{bh} or the formation of a \ac{ns} remnant~\cite{Breschi:2019srl} 
(see Appendix~\ref{app:inf:bajes} for details on \ac{bf} calculation).
An alternative method to assess the occurrence of \ac{pc} is to estimate 
the collapse time, $t_c$, to a \ac{bh}, which is a parameter introduced in \model~waveforms.
For practical purposes we assume that a system promptly collapses to a \ac{bh} if $t_c<t_c^{\rm thr} = 2~{\rm ms}$.
This threshold just serves as an upper limit for the \ac{pc} timescale;
since a proper definition of \ac{pc} requires inspection of the bulk merger dynamics~\cite{Radice:2018ghv}.

\section{Results}
 \label{sec:results}

\subsection{Inspiral merger}
\label{sec:im}

\begin{figure}[t]
	\includegraphics[width=0.5\textwidth]{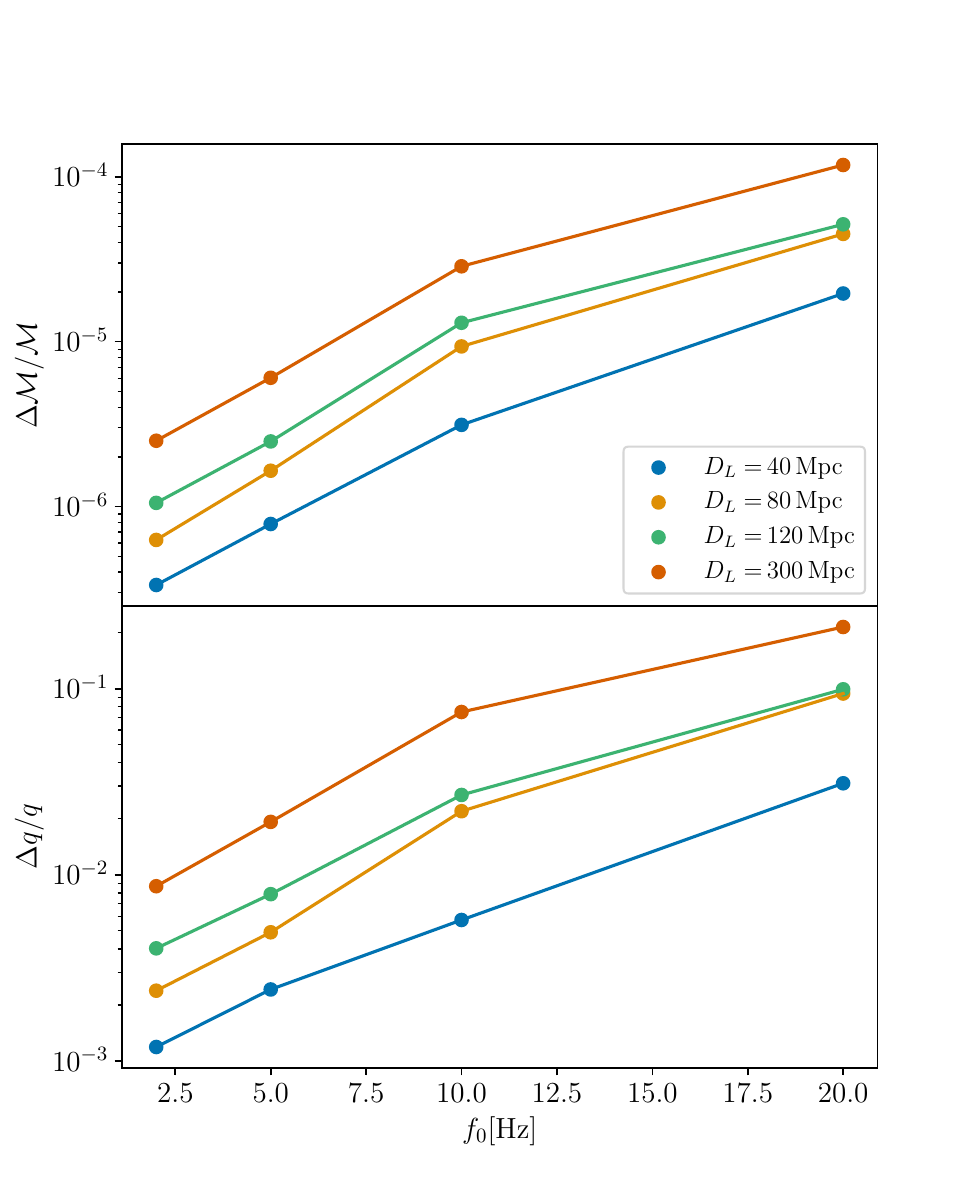}
    \caption{Relative precision on chirp mass (top) and mass ratio (bottom) with respect to 
	the different initial frequencies for a nonspinning \ac{bns} with $\Mc = 1.1976917\Msun$,
	$q = 1.5$, $\tilde{\Lambda} = 488$, at different luminosity distances.
	A lower initial frequency or a closer source leads to improved precision in 
	both the chirp mass and the mass ratio estimations.}
	 
	\label{fig:err_f0}
\end{figure}

\begin{figure*}[t]
    \includegraphics[width=0.99\textwidth]{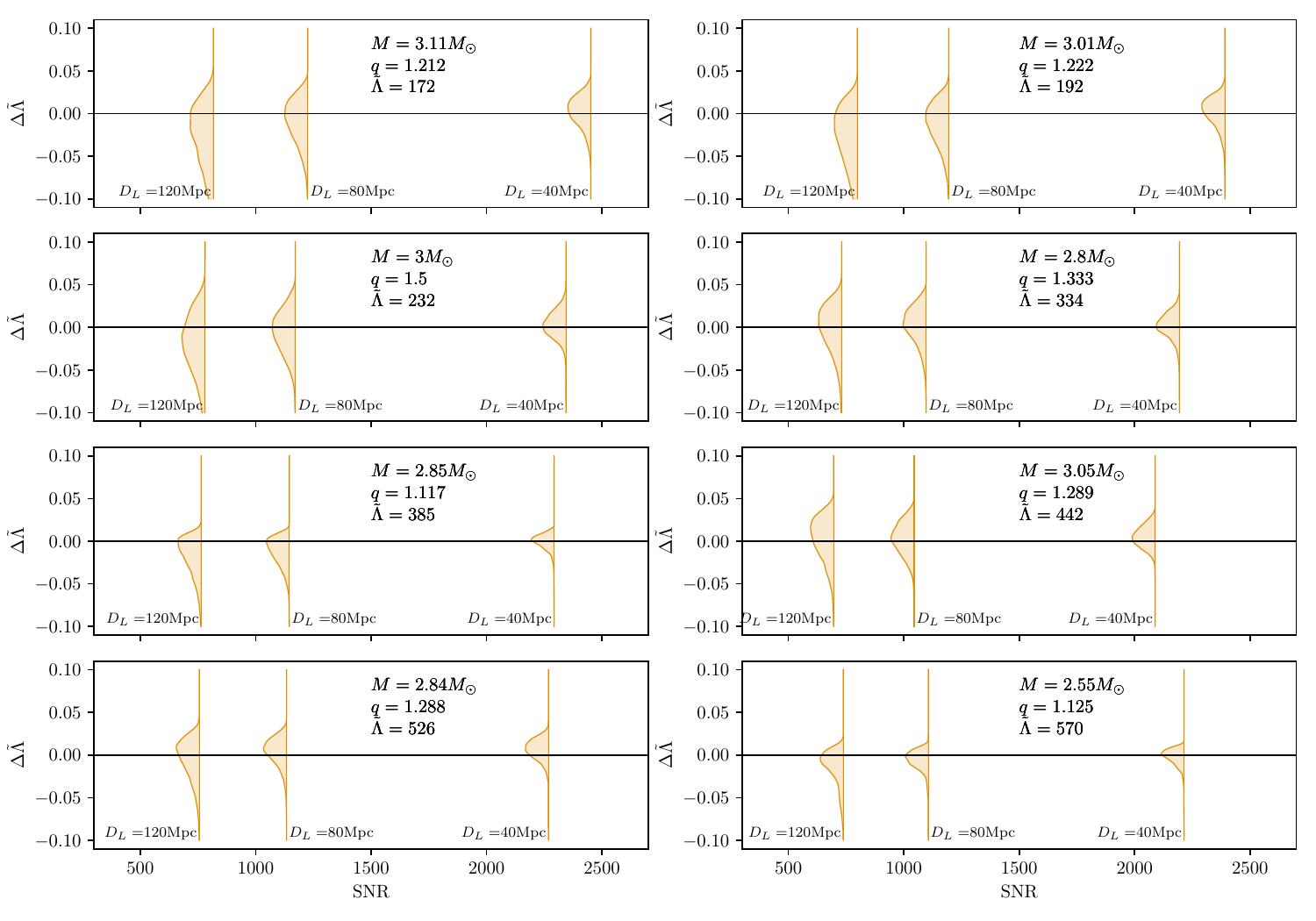}
    \caption{Posterior distributions for reduced tidal parameter $\tilde{\Lambda}$ with respect 
	to \ac{snr}, varying the injected luminosity distances, for eight fiducial binaries.
	The initial frequency is kept fixed to $f_0 = 5~{\rm Hz}$.
	As the \ac{snr} increases, \ie~the luminosity distance decreases, the width of the 
	distribution of $\tilde{\Lambda}$ shrinks.}
	\label{fig:lambda_SNR_dl}
\end{figure*}

\begin{figure*}[t]
    \includegraphics[width=0.99\textwidth]{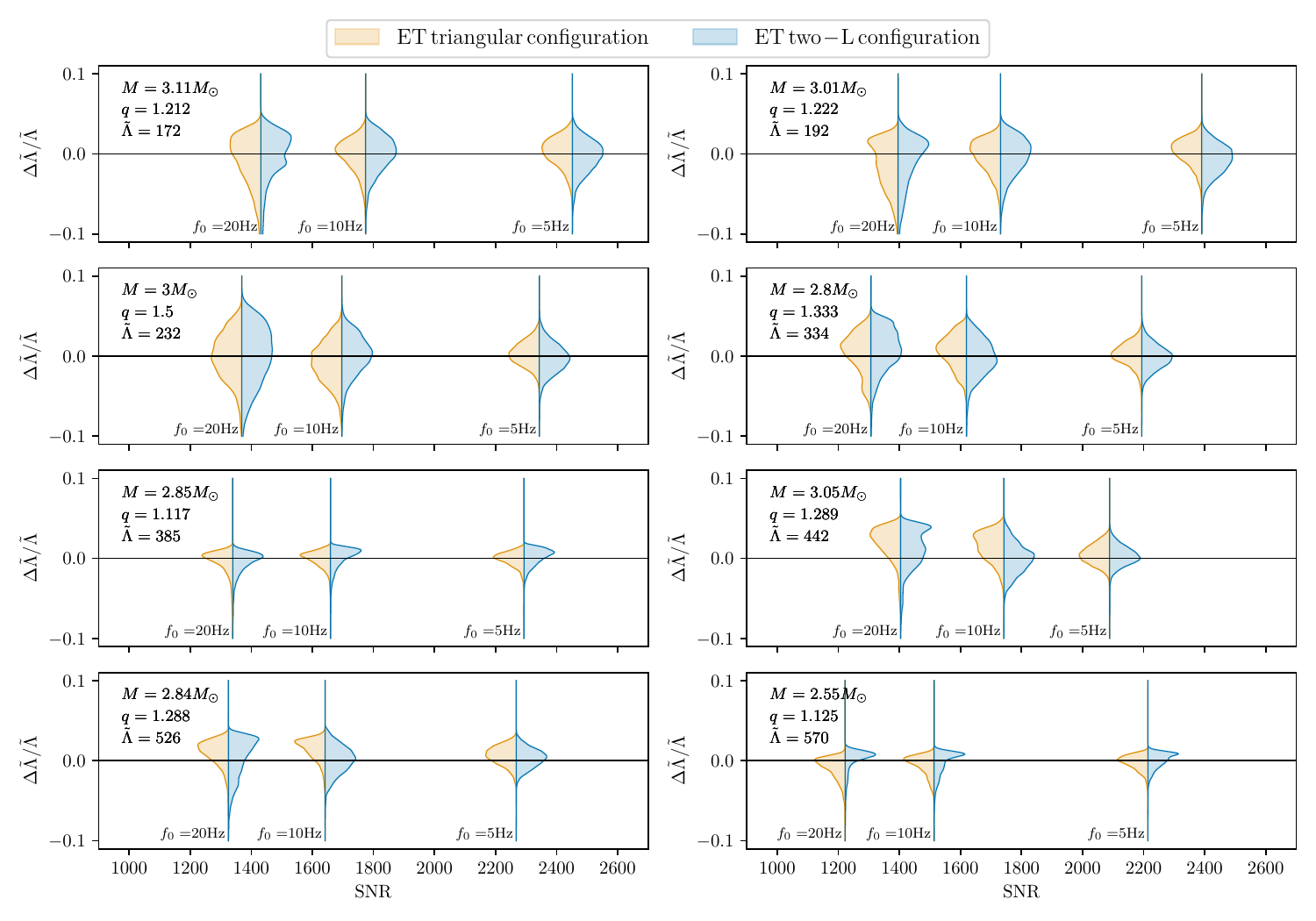}
    \caption{Posterior distributions for reduced tidal parameter $\tilde{\Lambda}$ with respect to 
	\ac{snr}, varying the initial frequencies of the detection, for eight fiducial binaries. 
	The luminosity distance is kept fixed to $D_L=40~{\rm Mpc}$. As the \ac{snr} increases, \ie~the 
	initial frequency decreases, the width of the distribution 	of $\tilde{\Lambda}$ shrinks. 
	Two different configurations of \ac{et}, triangular~\cite{Hild:2010id,Hild:2011np} in orange and two-L~\cite{Branchesi:2023mws} in blue, 
	give comparable results.} 
	\label{fig:lambda_SNR_f}
\end{figure*}

\begin{figure}[t]
    \includegraphics[width=0.45\textwidth]{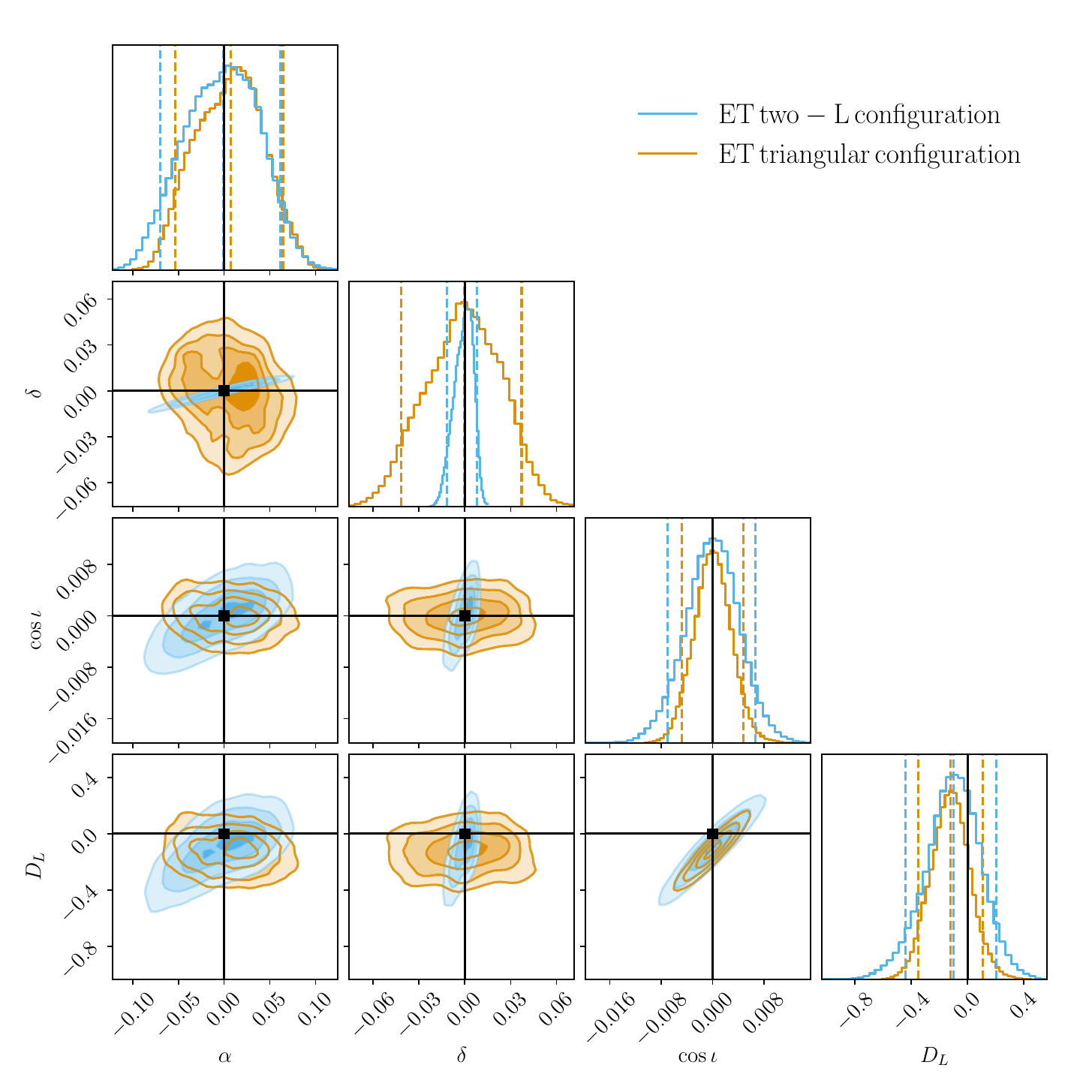}
    \caption{Comparison of the posterior distributions for extrinsic parameters
	for two different \ac{et} configurations: triangular~\cite{Hild:2010id,Hild:2011np} in orange 
	and two-L~\cite{Branchesi:2023mws} in 
	blue. The injections have been performed with {\teobspa}~template with 
	$\Mc = 1.344787\Msun$, $q = 1.21$, $\tilde{\Lambda} = 172$, 
	$D_L = 40\,{\rm Mpc}$ and $\iota =45^\circ$. The initial frequency $f_0$ 
	is $10\,{\rm Hz}$. We are showing the differences between the recovered 
	parameters and the injected ones, since the sky position
	corresponds to the relative best location for each detector.
	The posterior distributions are comparable, except for the estimation of
	the declination, where the two-L configuration provides tighter constraints.}
	\label{fig:extrinsic}
\end{figure}

\begin{table*}
	\centering	
	\caption{Summary of the \acp{pe} with {\teobspa} for a \ac{bns} with $\Mc = 1.1976917\Msun$, $q=1.5$, $\tilde{\Lambda}=488$, $\delta\tilde{\Lambda}=-60$, $\chi_{\rm eff} = 0$, $\iota = 0$.
	The first three columns report the injected luminosity distance, the initial frequency
	of the analyses and the \ac{snr} of the injected data; the last seven columns report the recovered values, with the
	median of the posterior distributions and the $90\%$ \acp{cl}.}
    \begin{tabular}{ccc|ccccccc}
		\hline
		\hline
		\multicolumn{3}{c|}{Injected properties}&\multicolumn{5}{c}{Recovered values}\\
		\hline
		$D_L~[{\rm Mpc}]$ & $f_{0}{[\rm Hz]}$ & SNR & $\Mc~[\Msun]$ & $q$ & $\tilde{\Lambda}$ & $\delta\tilde{\Lambda}$ & $\chi_{\rm eff}$ & $D_L~[{\rm Mpc}]$ & $\iota$\\
		\hline
		& 2 & 2261 & $1.1976917^{+0.0000002}_{-0.0000002}$ & $1.500^{+0.002}_{-0.002}$ & $490^{+9}_{-9}$ & $-50^{+30}_{-30}$& $-0.0001^{+0.0001}_{-0.0001}$ & $38.5^{+1.0}_{-0.8}$ & $0.26^{+0.07}_{-0.13}$ \\
		40& 5 & 2228 & $1.1976917^{+0.0000005}_{-0.0000005}$ & $1.500^{+0.004}_{-0.004}$ & $488^{+10}_{-10}$ & $-70^{+20}_{-20}$& $0.0000^{+0.0002}_{-0.0002}$ & $40.00^{+0.01}_{-0.01}$ & $0.02^{+0.01}_{-0.01}$ \\
		& 10 & 1613 & $1.197691^{+0.000002}_{-0.000002}$ & $1.498^{+0.011}_{-0.009}$ & $490^{+12}_{-14}$ & $-60^{+30}_{-30}$& $-0.0001^{+0.0009}_{-0.0007}$ & $40.01^{+0.02}_{-0.01}$ & $0.02^{+0.01}_{-0.01}$ \\
		& 20 & 1301 & $1.19769^{+0.00001}_{-0.00001}$ & $1.50^{+0.04}_{-0.05}$ & $490^{+30}_{-20}$ & $-60^{+70}_{-30}$& $-0.0002^{+0.0030}_{-0.0040}$ & $37.7^{+0.9}_{-0.8}$ & $0.33^{+0.06}_{-0.08}$ \\
		\hline
		& 2 & 1130 & $1.1976917^{+0.0000004}_{-0.0000004}$ & $1.500^{+0.003}_{-0.004}$ & $490^{+20}_{-20}$ & $-40^{+65}_{-40}$& $-0.0001^{+0.0002}_{-0.0003}$ & $76^{+2}_{-3}$ & $0.30^{+0.09}_{-0.13}$ \\
		80& 5 & 1114 & $1.1976915^{+0.0000010}_{-0.0000010}$ & $1.499^{+0.008}_{-0.007}$ & $490^{+30}_{-20}$ & $-40^{+80}_{-50}$& $-0.0001^{+0.0005}_{-0.0005}$ & $76^{+3}_{-2}$ & $0.31^{+0.09}_{-0.14}$ \\
		& 10 & 806 & $1.197690^{+0.000005}_{-0.000006}$ & $1.49^{+0.03}_{-0.03}$ & $500^{+50}_{-30}$ & $-40^{160}_{-50}$ & $-0.0005^{+0.0020}_{-0.0020}$ & $75^{+3}_{-3}$ & $0.33^{+0.09}_{-0.14}$ \\
		& 20 & 650 & $1.19769^{+0.00002}_{-0.00003}$ & $1.49^{+0.06}_{-0.14}$ & $500^{+90}_{-40}$ & $-30^{+220}_{-60}$ & $-0.0007^{+0.0060}_{-0.0110}$ & $75^{+3.0}_{-3}$ & $0.36^{+0.09}_{-0.13}$ \\
		\hline
		& 2 & 754 & $1.1976916^{+0.0000006}_{-0.0000006}$ & $1.498^{+0.006}_{-0.006}$ & $500^{+30}_{-30}$ & $0^{+110}_{-90}$ & $-0.0000^{+0.0007}_{-0.0006}$ & $113^{+5}_{-4}$ & $0.3^{+0.1}_{-0.2}$ \\
		120 & 5 & 743 & $1.197692^{+0.000001}_{-0.000002}$ & $1.50^{+0.01}_{-0.01}$ & $490^{+40}_{-30}$ & $-30^{+150}_{-70}$ & $-0.0001^{+0.0008}_{-0.0010}$ & $113^{+4}_{-5}$ & $0.3^{+0.1}_{-0.1}$ \\
		& 10 & 538 & $1.197689^{+0.000008}_{-0.000008}$ & $1.49^{+0.04}_{-0.04}$ & $510^{+60}_{-50}$ & $0^{+170}_{-100}$& $-0.001^{+0.003}_{-0.003}$ & $112^{+6}_{-5}$ & $0.4^{+0.1}_{-0.2}$ \\
		& 20 & 434 & $1.19768^{+0.00002}_{-0.00004}$ & $1.46^{+0.09}_{-0.14}$ & $530^{+70}_{-70}$ & $20^{+200}_{-110}$& $-0.003^{+0.008}_{-0.011}$ & $111^{+6}_{-5}$ & $0.4^{+0.1}_{-0.2}$ \\
		\hline
		& 2 & 301 & $1.197692^{+0.000002}_{-0.000002}$ & $1.50^{+0.01}_{-0.01}$ & $510^{+60}_{-50}$ & $80^{+160}_{-160}$ & $-0.0004^{+0.0010}_{-0.0013}$ & $270^{+10}_{-10}$ & $0.4^{+0.1}_{-0.2}$ \\
		300 & 5 & 297 & $1.197691^{+0.000003}_{-0.000004}$ & $1.49^{+0.02}_{-0.03}$ & $530^{+50}_{-70}$ & $100^{130}_{-180}$ & $-0.0002^{+0.0020}_{-0.0020}$ & $280^{+20}_{-20}$ & $0.4^{+0.2}_{-0.2}$ \\
		& 10 & 215 & $1.19769^{+0.00001}_{-0.00002}$ & $1.49^{+0.07}_{-0.11}$ & $500^{+100}_{-70}$ & $20^{+210}_{-130}$ & $-0.0005^{+0.0050}_{-0.0090}$ & $270^{+20}_{-20}$ & $0.4^{+0.1}_{-0.3}$ \\
		& 20 & 173 & $1.19768^{+0.00008}_{-0.00007}$ & $1.4^{+0.2}_{-0.3}$ & $540^{+70}_{-110}$ & $100^{+120}_{-220}$ & $-0.005^{+0.021}_{-0.020}$ & $270^{+20}_{-20}$ & $0.4^{+0.2}_{-0.2}$ \\\hline
		\hline
	\end{tabular}
	\label{tab:IM}
	\end{table*}

In this section we consider nine injection-recovery \acp{pe} with {\teobspa}.
We vary the initial frequency of the analyses and the luminosity distance of the binaries, 
comparing the recovered parameters with the injected ones. This approach allows us to investigate
how constraints on various parameters depend on the different \acp{snr} of the signals.

We generate artificial data in the frequency domain, using the triangular, triple-interferometer 
\ac{et} detector~\cite{Hild:2008ng,Punturo:2010zz,Maggiore:2019uih},
with ET-D sensitivity~\cite{Hild:2010id,Hild:2011np} a sampling rate of $4096~{\rm Hz}$, and a duration of $128~{\rm s}$.
We tested also longer durations to ensure that the length of the signal does not have any effects on the results. 
For example, with a signal length of 512 s and initial frequency $5~{\rm Hz}$, the posteriors are compatible within 90\% 
confidence regions to the ones shown below, since both injection and recovery are in the frequency domain.
The sky location corresponds to the optimal location for the \ac{et} detector $\{\alpha=2.64, 
\delta=0.762\}$ and the binaries are oriented with $\iota=0$ and $\psi=0$.
The injections are performed in zero noise, including only the dominant (2,2) mode.
The injected strain is analyzed in the frequency domain from $f_0$ to $2048~{\rm Hz}$, where 
$f_0\in[2,5,10,20]~{\rm Hz}$.
We set the priors for chirp mass $\Mc\in[0.8,1.5]\Msun$, for mass ratio $q\in[1,2]$ and 
for the tidal polarizability parameters $\Lambda_{1,2}\in[0,3000]$. The spins are aligned with the orbital
angular momentum, with an isotropic prior 
$\chi_{1,2}\in[-0.05,0.05]$ for the $z-$component. Finally, for the luminosity distance
we employ volumetric priors $D_L\in[10, 150]~{\rm Mpc}$, 
except when the injected luminosity distance is $300~{\rm Mpc}$, where we use $D_L\in[200, 400]~{\rm Mpc}$.
We used a uniform distribution for all the priors and we marginalize over reference phase.
In principle it is possible to use priors for the tidal polarizability parameters that 
consider $\Lambda_1 \leq \Lambda_2$, as shown in Fig.~1 of~\cite{LIGOScientific:2018cki}. 
We estimate the impact of this possible choice by removing samples from our posterior that 
do not satisfy this condition, which is equivalent to imposing this condition as a
prior and repeating the analysis. 
We notice a mildly impact in the results of the inference of the polarizability 
tidal parameters, obtaining a smaller 90\% credible intervals for $\tilde{\Lambda}$
and $\delta\tilde{\Lambda}$. For instance, for the injection discussed in Table~\ref{tab:IM}, 
the 90\% credible interval for $\tilde{\Lambda}$ changes from [481,499] to [481,498] 
at $D_L=40$Mpc and $f_0=2$Hz and from [430,610] to [450,610] at $D_L=300$Mpc and 
$f_0=20$Hz when imposing $\Lambda_1 < \Lambda_2$. The area of the posterior distributions
of the tidal polarizability parameters of Figs.~\ref{fig:lambda_SNR_dl} and~\ref{fig:lambda_SNR_f} 
shrinks as well, the difference, though, is really slight and not even present in all the cases. 
Therefore, we employed agnostic priors on $\Lambda$ for the results shown.

\subsubsection{Source parameters}

Figure~\ref{fig:err_f0} shows the relative precision on chirp mass and mass ratio calculated as
\be
\label{eq:rel_dis}
\frac{\Delta x}{x} = \frac{x_{95\%} - x_{5\%}}{x_{m}},
\ee
where $x_{m}$ is the median value and $x_{95\%}$ ($x_{5\%}$) corresponds to the $95$th ($5$th) percentile of the 
posterior distribution of the parameter $x$. 
The injected \ac{gw} signals correspond to nonspinning \ac{bns} with $\Mc = 1.1976917\Msun$, $q=1.5$, $\tilde{\Lambda}=488$ and $\delta\tilde{\Lambda}=-60$. 
These values are compatible with GW170817 event~\cite{LIGOScientific:2018mvr}. We vary the luminosity distance in $[40,80,120,300]~{\rm Mpc}$. 
The precision on both $\Mc$ and $q$ increases as the distance of the source decreases, since the
\ac{snr} is higher. Similarly, keeping a fixed luminosity distance, the precision increases
as the initial frequency $f_0$ of the analysis decreases, \eg~at $D_L = 300~{\rm Mpc}$, 
decreasing the frequency from $20~{\rm Hz}$ to $5~{\rm Hz}$, the precision on the measurement 
of both chirp mass and mass ratio improves of about one order of magnitude. 

Figure~\ref{fig:lambda_SNR_dl} shows the distributions of the differences between the 
recovered and the injected values of the reduced tidal parameter, 
\be
\frac{\Delta\tilde{\Lambda}}{\tilde{\Lambda}}= \frac{|\tilde{\Lambda}_{\rm rec} - \tilde{\Lambda}_{\rm inj}|}{\tilde{\Lambda}_{\rm inj}}\,,
\ee
with respect to the \ac{snr}, for eight different binaries.
As the \ac{snr} increases, that is when the luminosity distance decreases, the width of the distribution 
shrinks, meaning that the measurement of $\tilde{\Lambda}$ is more 
precise.
This is mostly due to the improvement in the chirp mass and mass ratio discussed above. 
We run the same test fixing the luminosity distance and varying the initial frequencies
of the analyses. The results are presented in Fig.~\ref{fig:lambda_SNR_f}. We obtain a similar 
result as before: as the \ac{snr} increases, namely as the initial frequency decreases, it is 
possible to have a better estimation on the value of $\tilde{\Lambda}$, \eg~of about a factor 
two, when the initial frequency changes from 20 to $5~{\rm Hz}$. 
Reducing the luminosity distance from $120~{\rm Mpc}$ to $40~{\rm Mpc}$, instead, improves the precision on the
reduced tidal parameters by a factor between two and three. At the same time, we notice that having a larger 
luminosity distance, \ie,~80~Mpc, but a smaller initial frequency, \ie,~5~Hz, enhances the
precision in the measurement of all the parameters with respect to a closer source ($D_L = 40~{\rm Mpc}$) 
at $f_0=20~{\rm Hz}$, even if the \ac{snr} is higher in the latter case. Thus, decreasing 
the initial frequency is an effective strategy to achieve better constraints in the estimation
of the binary parameters. 

Table~\ref{tab:IM} presents the recovered median values with the $90\%$ confidence regions
for the inferred parameters. In addition to the consideration above about the chirp mass and the mass
ratio, as the initial frequency decreases from 20~Hz to 2~Hz, the precision on the estimation of the reduced tidal 
parameter increases of about a factor two and the precision on the effective spin of one order of magnitude.
The precision of the luminosity distance and of the inclination, instead, is comparable for the 
different cases and the inferred values are not always compatible with the injected ones. This is due to the
known degeneracy between distance and inclination, which arises when considering waveforms
constructed only with the $(2,\pm 2)$ modes~\cite{Cutler:1994ys} Despite this, our results of the precision of the luminosity distance are consistent with
those reported in~\cite{Singh:2021bwn}. 

In Table~\ref{tab:IM} and in Table~\ref{tab:IM2} of Appendix~\ref{app:data} we also report the 
recovered values of the asymmetric tidal parameter 
$\delta\tilde{\Lambda}$. The estimation on this parameter carries substantial uncertainties, 
in some case larger than 100\%. The trend in the improvement of its precision 
follows the same behavior discussed above for the reduced tidal parameter. 
We notice that, when the \ac{snr} is sufficiently high, \ie~\ac{snr}$\gtrsim1500$ for $q>1.2$, 
the recovered values of $\delta\tilde{\Lambda}$ deviate significantly from zero, indicating a 
measurable asymmetry in the tidal polarizability parameters.
Furthermore, the joint estimation of $\delta\tilde{\Lambda}$ and $\tilde{\Lambda}$ leads to 
constraints on the individual tidal polarizability parameters. 
In few cases, when the mass ratio is $q\gtrsim1.289$ and the \ac{snr} has the highest value,
\ie~$f_0=$~5Hz and $D_L=$~40Mpc, we notice that both the posterior distributions of 
$\Lambda_1$ and $\Lambda_2$ are not compatible with zero, enabling us to confidently 
identify the source as a \ac{bns} merger.
However, in all the other cases, the posterior distribution of $\Lambda_1$ remains compatible with 
zero, preventing us from discarding the \ac{bh}-\ac{ns} hypothesis. 

\subsubsection{Comparison of ET configurations}

In Fig.~\ref{fig:lambda_SNR_f} we compare two different configurations of \ac{et}, 
the triangular one~\cite{Hild:2010id,Hild:2011np}, described in Sec.~\ref{sec:method:inf} and 
a two-L configuration~\cite{Branchesi:2023mws}, where the first detector is located in 
the Meuse-Rhine region and the second one in Sardinia, with the same sensitivity
as the triangular one~\cite{Hild:2010id,Hild:2011np} and a relative orientation between the two detectors of $45^\circ$.
We find that the posterior distributions of $\tilde{\Lambda}$ for the two 
configurations are comparable. Our results are in line with~\cite{Puecher:2023twf}.

We then perform a comparison between the two different configurations of \ac{et}
for what concerns the recovery of the extrinsic parameters, \ie~right ascension $\alpha$, declination 
$\delta$, inclination $\iota$ and luminosity distance $D_L$. Figure~\ref{fig:extrinsic} shows
the posterior distributions of the difference between the recovered and the injected values for 
the extrinsic parameters, since for each detector we use the best location. 
We notice that the $90\%$ confidence regions are comparable for the two configurations for all the 
parameters, except for the declination, which is better recovered in the case of the two-L configuration. 

\citet{Branchesi:2023mws,Begnoni:2025oyd}, and \citet{Puecher:2023twf} compared various \ac{et} geometries 
using a Fisher matrix approach in the two former cases and a Bayesian \ac{pe} in the latter two. 
They all concluded that an improvement in the precision on the inference of the source 
parameters comes primarily from the arms length, while changing shape or alignment of the detectors plays 
a minimal role in the accuracy increase. Additionally,~\citet{Branchesi:2023mws}
found that the estimation of the source parameters is comparable for the two configurations we 
considered, while the triangular performs better in the estimation of the inclination angle. 
Regarding the angular localization, their results indicate that the two-L is better compared to 
the triangular one. This improvement is attributed to the ability of the two-L setup to 
partially triangulate, thanks to its longer baseline. These findings
are compatible with our results.

\subsection{Inspiral-merger-postmerger}
\label{sec:impm}

\begin{table*}
	\centering	
	\caption{Summary of the \acp{pe} with {\teobspa} for \ac{im} and {\teobnrpmw} for \ac{impm} case.
	The first five columns report the injected parameters, included the \ac{eos} used; the central six columns report the recovered values, with the
	median of the posterior distributions and the $90\%$ \acp{cl} and the last column displays
	the \acp{bf}, with statistical uncertainties as in~\cite{Skilling:2006gxv}.}
	\begin{tabular}{ccccc|ccc|ccc|c}
			\hline
			\hline
			\multicolumn{5}{c|}{Injected properties}&\multicolumn{3}{c|}{Recovered values IM}&\multicolumn{3}{c|}{Recovered values IMPM}&\multicolumn{1}{c}{}\\
			\hline
			morph. & $\Mc~[\Msun]$ & $q$ & $\tilde{\Lambda}$ & SNR & $\Mc~[\Msun]$ & $q$ & $\tilde{\Lambda}$ & $\Mc~[\Msun]$ & $q$ & $\tilde{\Lambda}$ & $\log\mathcal{B}_{\rm{IM}}^{\rm{IMPM}}$\\
			\hline
			DD2 & 1.339520 & 1 & 403 & 1427 & $1.339519^{+0.000001}_{-0.000001}$ & $1.013^{+0.015}_{-0.012}$ & $400^{+4}_{-6}$  & $1.339519^{+0.000001}_{-0.000001}$ & $1.016^{+0.012}_{-0.011}$ & $400^{+4}_{-6}$ & $32.29^{+0.47}_{-0.47}$\\
		    BHB$\Lambda\phi$ & 1.319184 & 1 & 434 & 1409 & $1.319184^{+0.000001}_{-0.000001}$ & $1.013^{+0.014}_{-0.012}$ & $431^{+4}_{-6}$ &  $1.319184^{+0.000001}_{-0.000001}$ & $1.016^{+0.011}_{-0.011}$ & $431^{+5}_{-6}$ & $31.98^{+0.47}_{-0.47}$\\
			DD2 & 1.522426 & 1 & 178 & 1583 & $1.522425^{+0.000001}_{-0.000001}$ & $1.011^{+0.013}_{-0.010}$ & $177^{+2}_{-3}$  & $1.522425^{+0.000001}_{-0.000002}$ & $1.016^{+0.012}_{-0.011}$ & $176^{+3}_{-3}$ & $29.54^{+0.48}_{-0.48}$\\
			LS220 & 1.170202 & 1 & 703 & 1278 & $1.170202^{+0.000001}_{-0.000001}$ & $1.015^{+0.014}_{-0.013}$ & $699^{+6}_{-8}$ & $1.170202^{+0.000001}_{-0.000001}$ & $1.015^{+0.012}_{-0.012}$ & $699^{+7}_{-8}$ & $24.99^{+0.47}_{-0.47}$\\
			DD2 & 1.315304 & 1.289 & 442 & 1405 & $1.315304^{+0.000002}_{-0.000002}$ & $1.289^{+0.002}_{-0.002}$ & $443^{+12}_{-14}$ & $1.315304^{+0.000002}_{-0.000002}$ & $1.290^{+0.002}_{-0.002}$ & $440^{+12}_{-13}$ & $22.25^{+0.48}_{-0.48}$\\
			SLy & 1.175243 & 1 & 394 & 1466 & $1.175243^{+0.000001}_{-0.000001}$ & $1.014^{+0.014}_{-0.012}$ & $390^{+5}_{-6}$ & $1.175243^{+0.000001}_{-0.000001}$ & $1.014^{+0.011}_{-0.011}$ & $392^{+4}_{-5}$ & $19.39^{+0.48}_{-0.48}$\\
			BLh & 1.227164 & 1 & 412 & 1329 & $1.227163^{+0.000001}_{-0.000001}$ & $1.012^{+0.015}_{-0.011}$ & $409^{+5}_{-6}$ & $1.227163^{+0.000001}_{-0.000001}$ & $1.017^{+0.012}_{-0.012}$ & $408^{+4}_{-6}$ & $18.90^{+0.47}_{-0.47}$\\
			DD2 & 1.043690 & 1 & 1638 & 1164 & $1.043690^{+0.000000}_{-0.000001}$ & $1.015^{+0.016}_{-0.013}$ & $1631^{+9}_{-18}$ & $1.043690^{+0.000001}_{-0.000001}$ & $1.018^{+0.014}_{-0.013}$ & $1629^{+10}_{-17}$ & $16.72^{+0.46}_{-0.46}$\\
			BLh & 1.341381 & 1 & 222 & 1429 & $1.341381^{+0.000001}_{-0.000001}$ & $1.012^{+0.013}_{-0.011}$ & $220^{+3}_{-4}$  & $1.341380^{+0.000001}_{-0.000001}$ & $1.013^{+0.012}_{-0.010}$ & $221^{+3}_{-4}$ & $15.16^{+0.48}_{-0.48}$\\
			SFHo & 1.130733 & 1 & 538 & 1243 & $1.130733^{+0.000000}_{-0.000001}$ & $1.014^{+0.015}_{-0.012}$ & $534^{+6}_{-8}$ & $1.130732^{+0.000001}_{-0.000001}$ & $1.017^{+0.012}_{-0.012}$ & $534^{+6}_{-8}$ & $15.05^{+0.46}_{-0.46}$\\
			BLh & 1.239566 & 1 & 385 & 1339 & $1.239565^{+0.000001}_{-0.000001}$ & $1.117^{+0.004}_{-0.004}$ & $385^{+7}_{-12}$ & $1.239565^{+0.000001}_{-0.000001}$ & $1.117^{+0.004}_{-0.003}$ & $385^{+6}_{-9}$ & $14.36^{+0.47}_{-0.47}$\\
			SFHo & 1.273413 & 1 & 249 & 1369 & $1.273412^{+0.000001}_{-0.000001}$ & $1.013^{+0.014}_{-0.011}$ & $247^{+3}_{-4}$ & $1.273412^{+0.000001}_{-0.000001}$ & $1.016^{+0.013}_{-0.012}$ & $246^{+4}_{-5}$ & $14.19^{+0.47}_{-0.47}$\\
			SFHo & 1.344787 & 1.212 & 172 & 1432 & $1.344787^{+0.000002}_{-0.000002}$ & $1.213^{+0.002}_{-0.002}$ & $171^{+6}_{-9}$  & $1.344787^{+0.000002}_{-0.000001}$ & $1.213^{+0.002}_{-0.002}$ & $171^{+5}_{-7}$ & $13.94^{+0.48}_{-0.48}$\\	
			SLy & 1.058974 & 1 & 743 & 1178 & $1.058974^{+0.000000}_{-0.000001}$ & $1.013^{+0.015}_{-0.012}$ & $738^{+7}_{-10}$ & $1.058974^{+0.000001}_{-0.000001}$ & $1.018^{+0.012}_{-0.013}$ & $737^{+7}_{-10}$ & $12.99^{+0.46}_{-0.46}$\\
			SLy & 1.244295 & 1 & 264 & 1344 & $1.244295^{+0.000001}_{-0.000001}$ & $1.012^{+0.014}_{-0.011}$ & $262^{+4}_{-4}$ & $1.244295^{+0.000001}_{-0.000001}$ & $1.016^{+0.013}_{-0.012}$ & $261^{+4}_{-5}$ & $12.89^{+0.47}_{-0.47}$\\
			SLy & 1.303371 & 1 & 191 & 1396 & $1.303371^{+0.000001}_{-0.000001}$ & $1.012^{+0.013}_{-0.011}$ & $189^{+3}_{-4}$  & $1.303370^{+0.000001}_{-0.000001}$ & $1.015^{+0.012}_{-0.011}$ & $189^{+3}_{-4}$  & $12.51^{+0.47}_{-0.47}$\\
			LS220 & 1.223692 & 1.287 & 525 & 1325 & $1.223692^{+0.000001}_{-0.000002}$ & $1.288^{+0.002}_{-0.002}$ & $525^{+13}_{-16}$ & $1.223692^{+0.000001}_{-0.000001}$ & $1.288^{+0.002}_{-0.002}$ & $525^{+11}_{-15}$ & $12.22^{+0.47}_{-0.47}$\\
			SLy & 1.304020 & 1.222 & 192 & 1396 & $1.304020^{+0.000001}_{-0.000002}$ & $1.220^{+0.002}_{-0.002}$ & $192^{+6}_{-9}$  & $1.304020^{+0.000001}_{-0.000001}$ & $1.220^{+0.002}_{-0.002}$ & $192^{+5}_{-7}$ & $9.78^{+0.48}_{-0.48}$\\
			SLy & 1.183075 & 1.257 & 365 & 1289 & $1.183075^{+0.000001}_{-0.000001}$ & $1.257^{+0.002}_{-0.002}$ & $365^{+10}_{-14}$ & $1.183075^{+0.000001}_{-0.000001}$ & $1.257^{+0.002}_{-0.002}$ & $363^{+11}_{-14}$ & $8.91^{+0.47}_{-0.47}$\\
			SFHo & 1.339422 & 1 & 130 & 1427 & $1.339422^{+0.000001}_{-0.000001}$ & $1.012^{+0.013}_{-0.011}$ & $129^{+3}_{-3}$  & $1.339422^{+0.000001}_{-0.000001}$ & $1.014^{+0.012}_{-0.011}$ & $129^{+3}_{-3}$ & $7.17^{+0.48}_{-0.48}$\\
									\hline		
			2B & 1.175243 & 1 & 126 & 1283 & $1.175243^{+0.000001}_{-0.000001}$ & $1.013^{+0.013}_{-0.011}$ & $124^{+4}_{-4}$ & $1.175243^{+0.000001}_{-0.000001}$ & $1.013^{+0.013}_{-0.011}$ & $124^{+4}_{-4}$ & $-3.63^{+0.47}_{-0.47}$\\
									DD2 & 1.522426 & 1 & 178 & 1583 & $1.522425^{+0.000001}_{-0.000002}$ & $1.015^{+0.013}_{-0.011}$ & $177^{+2}_{-3}$ & $1.522425^{+0.000001}_{-0.000002}$ & $1.015^{+0.013}_{-0.011}$ & $177^{+2}_{-3}$  & $-5.84^{+0.47}_{-0.47}$\\
			SLy & 1.303371 & 1 & 191 & 1396 & $1.303370^{+0.000001}_{-0.000001}$ & $1.014^{+0.013}_{-0.011}$ & $189^{+3}_{-4}$ & $1.303370^{+0.000001}_{-0.000001}$ & $1.014^{+0.013}_{-0.011}$ & $189^{+3}_{-4}$ & $-6.28^{+0.47}_{-0.47}$\\
			\hline
			\hline
	\end{tabular}
	\label{tab:IMPM}
\end{table*}

\begin{figure}[t]
	\includegraphics[width=0.45\textwidth]{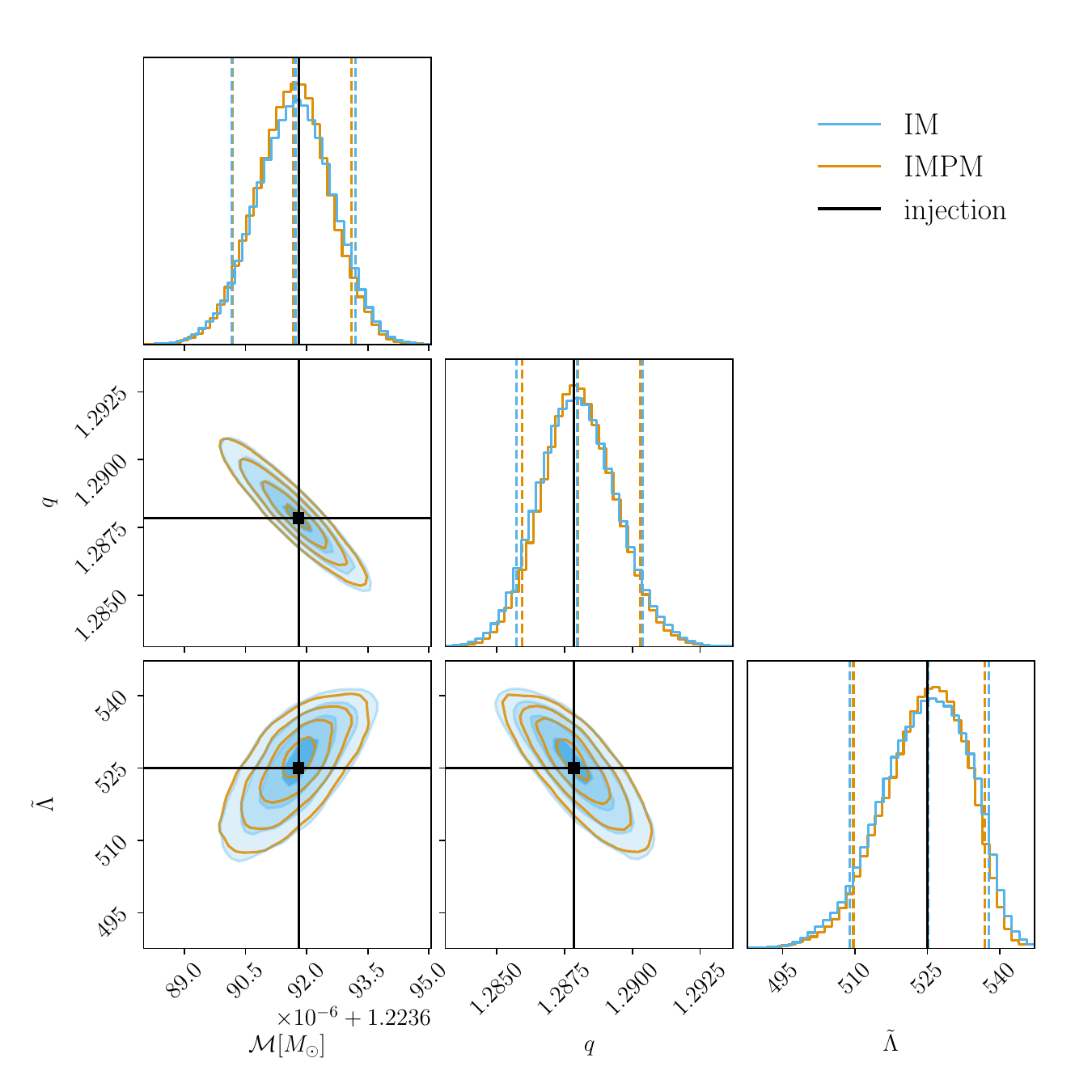}
	\caption{Posterior distributions for intrinsic parameters recovered from the injection of a \ac{bns}
	using LS220 \ac{eos} ($\Mc = 1.223692\,\Msun$), as in Table~\ref{tab:IMPM}. The recovery is performed both 
	with {\teobspa} (\ac{im}) and {\teobnrpmw} (\ac{impm}). The injected values are in black.
	The posterior distributions obtained with the two models are analogous, since most of the information is 
	included in the \ac{im} section.}
	\label{fig:IM_IMPM}
\end{figure}

\begin{figure*}[t]
	\includegraphics[width=0.99\textwidth]{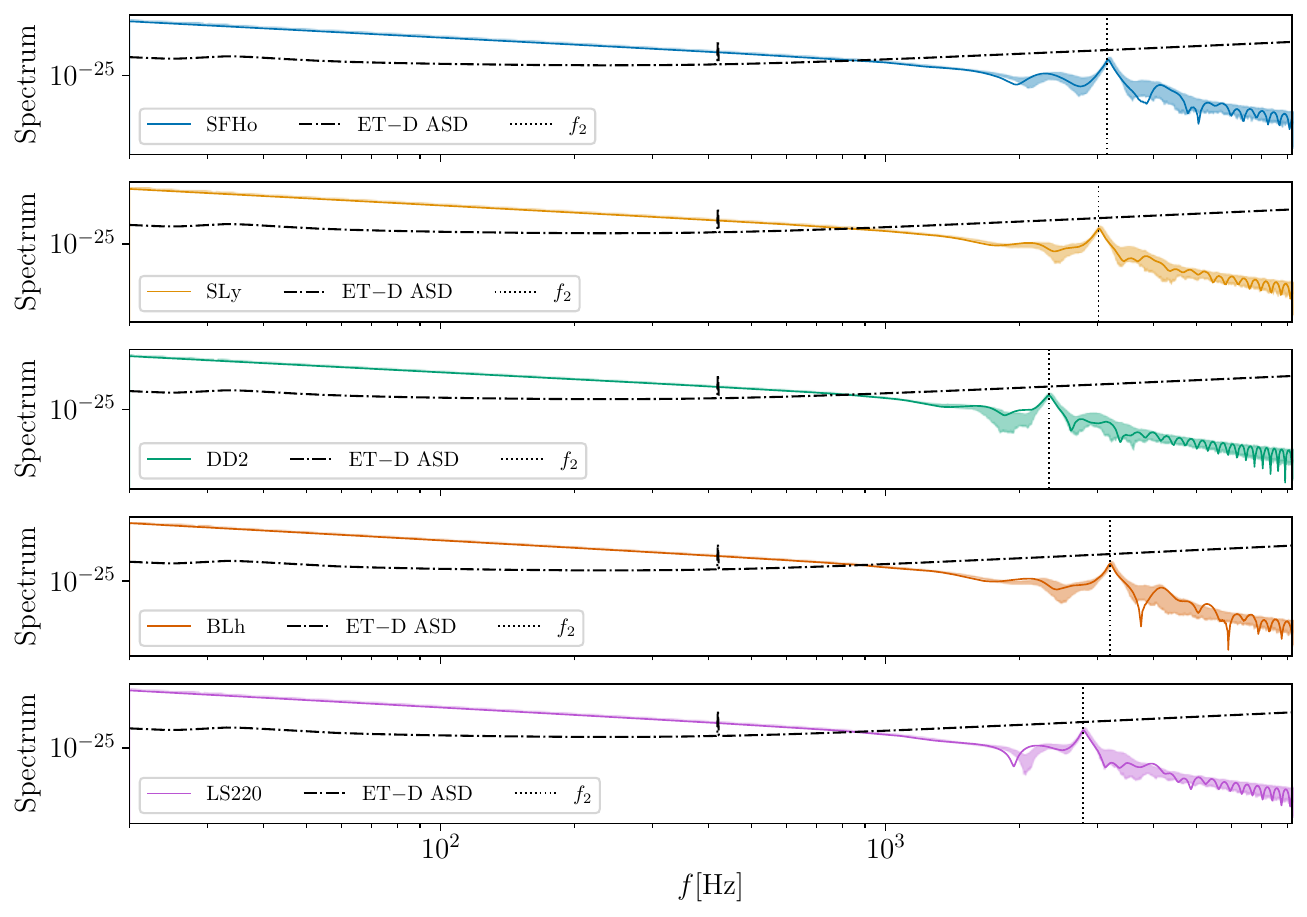}
	\caption{Injected and recovered spectra of five exemplary binaries, with different \acp{eos}.
	The solid colored lines report the injected waveforms, the shadowed regions correspond 
	to the $90\%$ credibility regions and the vertical black dotted lines represent
	the \ac{pm} peaks at $f_2$. We also report the ASD of the ET detector~\cite{Hild:2010id,Hild:2011np}.
	Both the inspiral part and the peak of the \ac{pm} are well recovered in all the cases.}
	\label{fig:reconstr_spectrum}
\end{figure*}

This section is dedicated to the analysis of \ac{impm} injection-recovery \acp{pe}. We 
employed the complete {\teobnrpmw}~template for the injections of \acp{gw} from different binaries, 
\ie~with different \acp{eos}.
As in the \ac{im}-only case, the detector used is \ac{et}, with triangular shape and ET-D sensitivity~\cite{Hild:2010id,Hild:2011np}.
The sky location is the same as in~\ref{sec:im} and the luminosity distance is $40~{\rm Mpc}$. The sampling 
rate is $16384~{\rm Hz}$ and the signal length $128~\rm{s}$,
the analyses are performed in the frequency range $f\in[20,8192]~\rm{Hz}$.
The priors for chirp mass 
are $\Mc\in[0.8,2]\,\Msun$, for mass ratio $q\in[1,4]$, for the tidal polarizability 
parameters $\Lambda_{1,2}\in[0,4000]$ and for the 
luminosity distance are volumetric $D_L\in[10, 1000]~{\rm Mpc}$. We perform the inference 
under the assumption of nonspinning binaries and we keep all the other extrinsic 
parameters fixed to the injected values. We add the recalibration parameters 
$\recalib_{\rm fit}$ to account for intrinsic errors of the calibrated formulas (see paper I).

Figure~\ref{fig:IM_IMPM} displays an example of posterior distributions for the LS220  
binary of Table~\ref{tab:IMPM} with $\Mc = 1.223692\,\Msun$. We compare between \ac{im} and \ac{impm} recoveries. 
The posteriors distributions are, in both cases, correctly centered on the injected values (black lines) and they are similar.
We observe that the inclusion of the \ac{pm} does not significantly improve the constraints 
on chirp mass, mass ratio and tidal polarizability parameter compared to the \ac{im}-only \ac{pe}. 
This outcome is primarily due to the dominant contribution of the \ac{im} segment, which exhibits 
a \ac{snr} on the order of $10^3$, whereas the \ac{snr} of the \ac{pm} is only on the order of 
ten (the \ac{snr} threshold for detectability of the \ac{pm} is 7 [paper II]). Nevertheless, having a model 
that incorporates the \ac{pm} remains essential for 
distinguishing between the characteristic \ac{pm} features of a \ac{ns} remnant and the case of a \ac{pc}, 
by using \acp{bf} for model selection, as discussed below.

Figure~\ref{fig:reconstr_spectrum} presents five illustrative recovered spectra, 
in which the injected waveforms are included within the $90\%$ 
\acp{cl} of the reconstructed spectra. Particularly, both the inspiral part and the frequency
of the peak of the \ac{pm}, $f_2$, are well recovered with tight error bars for all the binaries, 
\ie~it is possible to obtain constraints on $f_2$ for different \acp{eos}
with an uncertainty smaller than 8\%.
The high-frequency features and the region between the merger and the peak of the \ac{pm} 
are consistent with the injection, but, in these cases, the $90\%$ \ac{cl} are broader, 
indicating lower precision on the detection compared to the inspiral segment. This reduced 
precision arises because the strain amplitude is about two orders of magnitude smaller 
and the detector sensitivity is lower in the high-frequency region. 

In Table~\ref{tab:IMPM} we report the results of the \ac{pe} analyses we performed.
In all the cases we use both the \ac{im} waveform {\teobspa} and the \ac{impm} waveform {\teobnrpmw}~for the recoveries. 
The inferred values for the two cases are comparable and always compatible with the injected values.
We compute the \acp{bf} as
\be
	\log \mathcal{B}_{\rm{IM}}^{\rm{IMPM}} = \log \frac{p(d|{\rm IMPM})}{p(d|{\rm IM})}\,, 
\ee 
where $p(d|H)$ is the evidence. 
In the upper part of the table the \acp{bf} are
larger than zero, meaning that the \ac{impm} model is preferred over the \ac{im} model.
All the $\log$\ac{bf} values are on the order of ten, whereas the associated uncertainties, estimated
with the criterion introduced in~\cite{Skilling:2006gxv}, are on the order $10^{-1}$. This 
indicates a high degree of statistical confidence, supporting a clear detection of the 
presence of the \ac{pm}.
In the bottom part of the table, instead, we test the ability of our pipeline to recover \ac{pc}.
The relative discussion is postponed to Sec.~\ref{sec:pc}.
In order to discard the possibility that the \ac{bf} of the \ac{impm} analyses is biased,
due to the fact that the \ac{impm} analyses may recover a higher \ac{snr} than the \ac{im}-only one, 
we check the recovered \ac{snr} values and compare the \ac{im}-only case with
the \ac{impm} one. Despite the values of the recovered \ac{snr} in the \ac{im}-only
case being slightly smaller, they are compatible within the 90\%
confidence range with the recovered ones from the \ac{impm} case and with
the injected ones as well. For example for the binary with SLy \ac{eos} and 
$\Mc = 1.058974M_\odot$ the injected \ac{snr} is 1177.978, the recovered 
\ac{snr} in the \ac{im}-only case is $1177.900^{+0.012}_{-0.084}$ and the recovered 
one for the \ac{impm} case is $1177.924^{+0.037}_{-0.075}$.
Therefore we consider that the \ac{snr} is not the main reason for the \ac{bf} to 
favor the \ac{impm} case. We also note that the \acp{snr} of the \ac{im} and the 
\ac{pm} are summed in quadrature and, since the \ac{snr} of the \ac{im} is two order 
of magnitude larger compared to the \ac{pm} one, adding the \ac{pm} will
not change the total \ac{snr} (almost) at all.

\subsection{EoS constraints}
\label{sec:highdens}

\begin{figure*}[t]
	\includegraphics[width=0.99\textwidth]{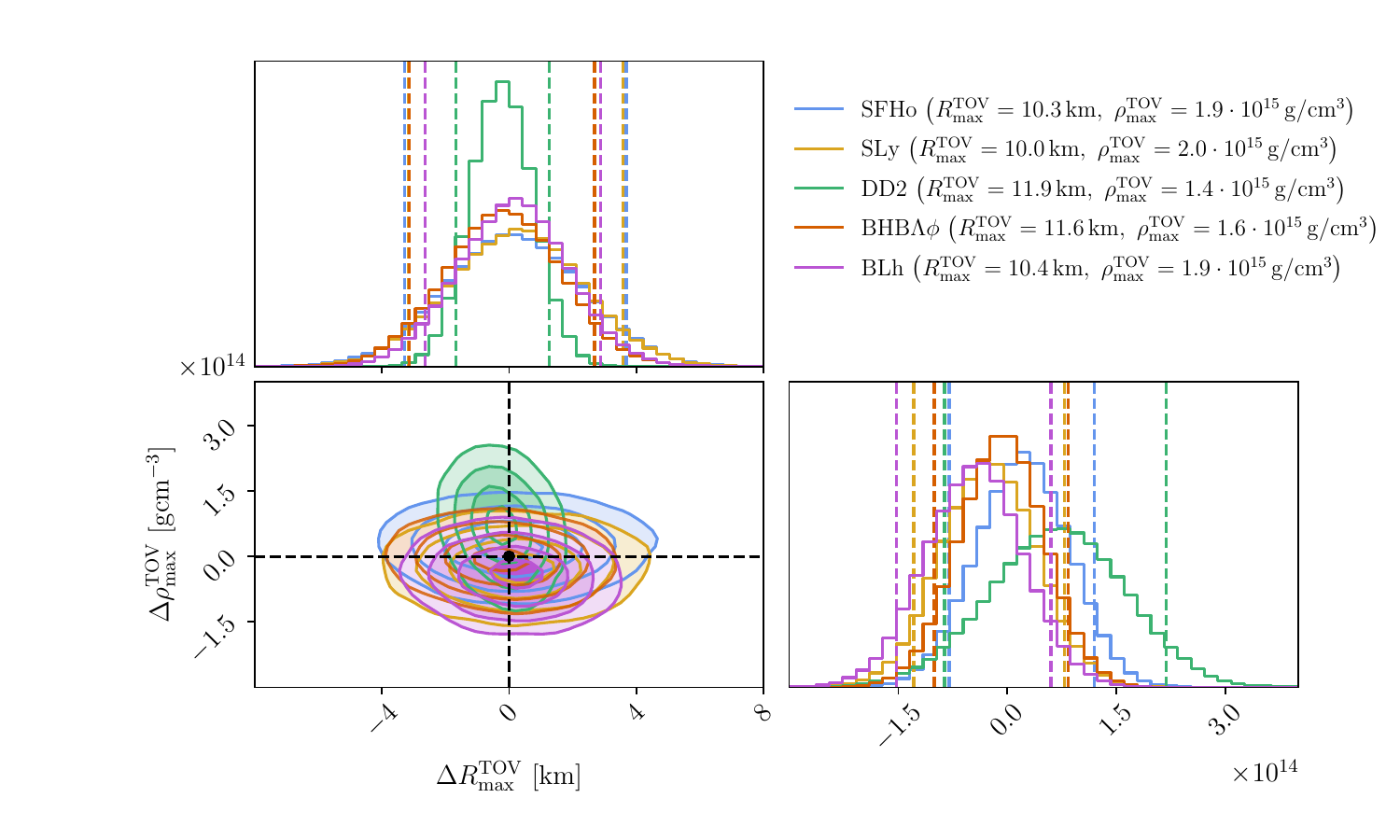}
	\caption{Posterior distributions of the difference between recovered and theoretical 
	values for maximum density $\rhomax$ and radius at the maximum mass $\Rmax$ for five different \acp{eos}.
	The recovered values are computed with~Eqs.~\eqref{eq:rhomax} and~\eqref{eq:Rmax} and 
	are consistent with the theoretical values from the \ac{eos} sequences, as reported in 
	the legend.}
	\label{fig:rhormax}
\end{figure*}

\begin{figure*}[t]
	\includegraphics[width=0.45\textwidth]{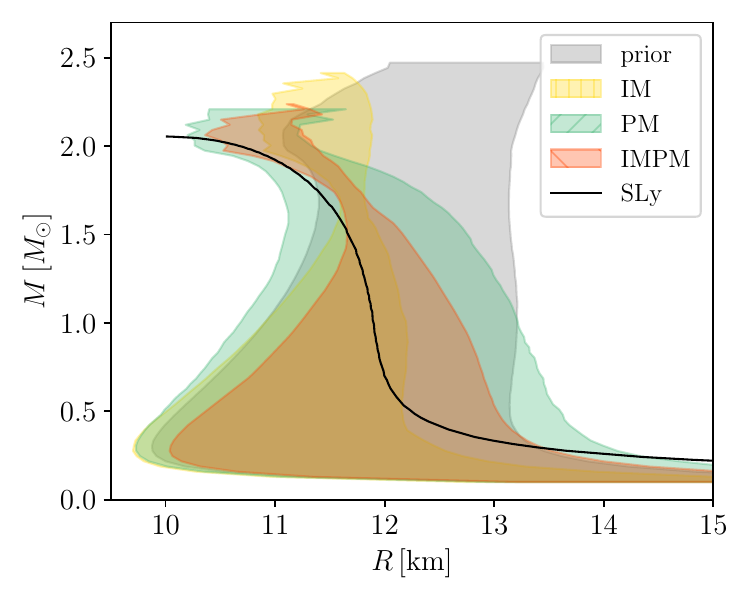}
	\includegraphics[width=0.45\textwidth]{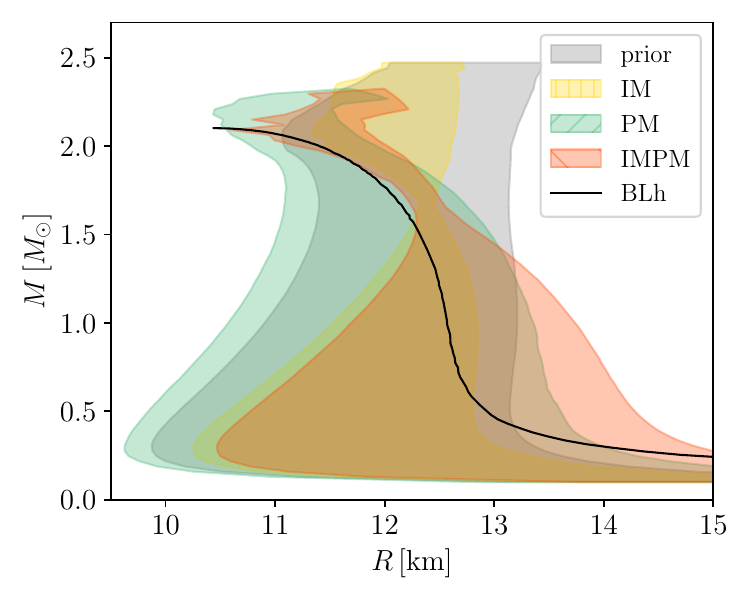}
	\caption{Mass-radius diagram constraints from two \ac{bns} observations, with
	different \acp{eos}: SLy ($\Mc=1.244295\Msun$) on the left and BLh ($\Mc=1.227164\Msun$) on the right
	(See Table~\ref{tab:IMPM}). 
	The gray area (prior) corresponds to the two millions \ac{eos} sample of~\cite{Godzieba:2020tjn}. 
	The colored areas are the 90$\%$ credibility regions given by \ac{im}, \ac{pm} and
	\ac{impm} inferences. The theoretical mass-radius curves are
	reported in black.
	The inclusion of the \ac{pm} gives information in the high-density regime. The further combination
	with the \ac{im} inference leads to tighter constraints of the mass-radius diagram.}
	\label{fig:eos_constraint}
\end{figure*}

\begin{figure}[t]
	\includegraphics[width=0.45\textwidth]{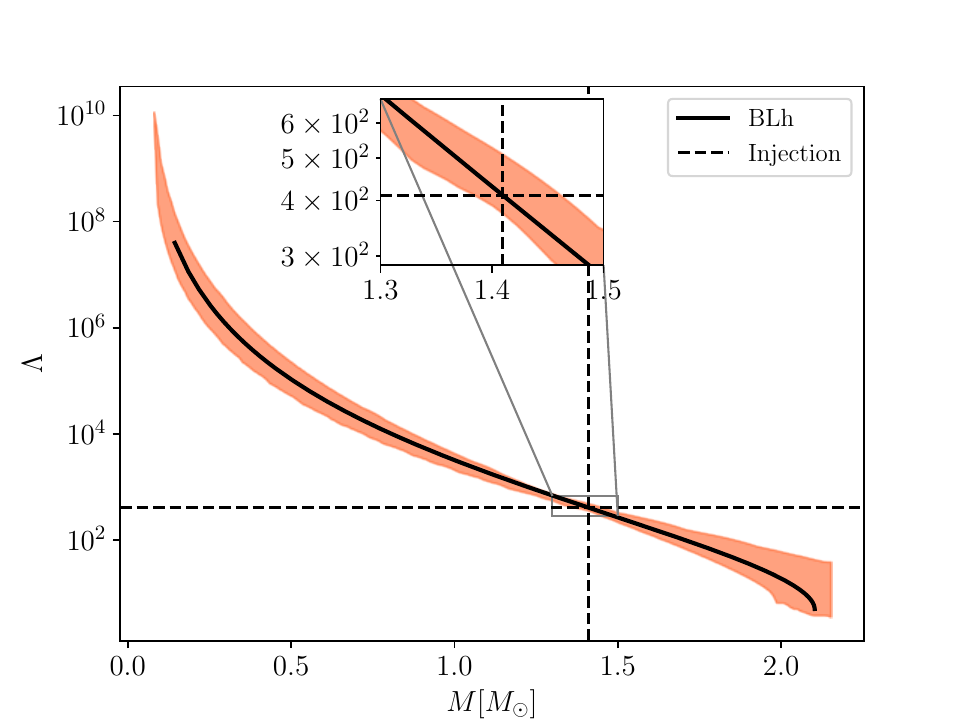}
	\caption{Tidal polarizability parameter as function of the mass for BLh ($\Mc=1.227164\Msun$) \ac{eos} 
	(See Table~\ref{tab:IMPM}).	The colored region represents the 90$\%$ \ac{cl} for the
	\ac{impm} inference. It is compatible and tightly constrained around the theoretical $\Lambda(M)$ curve,
	shown in black. The dashed lines correspond to the injected values.}
	\label{fig:lambda_m}
\end{figure}

\begin{figure}[t]
	\includegraphics[width=0.45\textwidth]{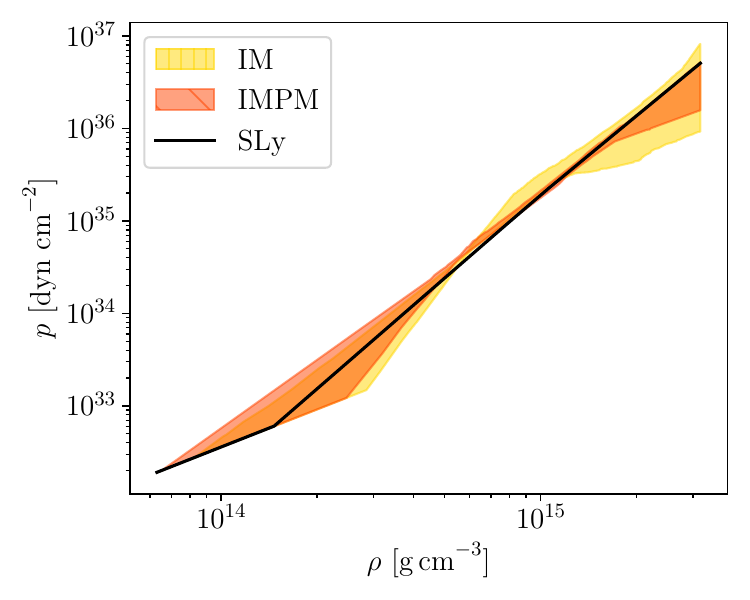}
	\caption{Pressure-density curve for SLy ($\Mc=1.244295\Msun$) \ac{eos} 
	(See Table~\ref{tab:IMPM}).	The colored regions represent the 90$\%$ \acp{cl} for the \ac{im} (yellow) and
	\ac{impm} (red) inferences. The theoretical $p(\rho)$ relation is reported in black. 
	The constrained areas are compatible with the theoretical predictions and the \ac{impm} analysis gives a 
	tighter constraint at higher densities.}
	\label{fig:p_rho}
\end{figure}

In this section we present the results of the \ac{eos} study enabled by our \ac{pe} analyses.
In particular \ac{eos} sampling can be performed starting from the results of 
the \ac{im}-only \acp{pe}, employing only the masses and the tidal polarizability parameters, 
or from the results of the \ac{pm}, incorporating also the constraints from the quasiuniversal relations for the 
maximum density $\rhomax$ and the radius at the maximum mass 
$\Rmax$, as described in Sec.~\ref{sec:method:nrrel}. 

We begin examining the distribution of the maximum density and the radius at the 
maximum mass, computed from the values of $f_2$ using the quasiuniversal relations~\eqref{eq:rhomax}
and~\eqref{eq:Rmax}, respectively. Figure~\ref{fig:rhormax} displays the posterior distributions
of $\Delta\rhomax = \rho_{\rm max}^{\rm TOV,PE}-\rhomax$ and $\Delta\Rmax = R_{\rm max}^{\rm TOV,PE}-\Rmax$, 
for five different \acp{eos}.
$\rhomax$ and $\Rmax$ corresponds to the maximal theoretical values 
obtained from the stable \ac{eos} sequence, while $\rho_{\rm max}^{\rm TOV,PE}$ and $R_{\rm max}^{\rm TOV,PE}$
are computed from the fits.
We observe that for soft \acp{eos} the distributions of $\Delta\rhomax$ are exactly centered 
around zero, whereas for stiffer \acp{eos}, such as DD2, the distributions are shifted away from zero, 
although they remain compatible with zero within the $90\%$ \ac{cl}. This discrepancy arises from 
the fit relations used in the analysis. In particular examining Fig.~1 of~\cite{Breschi:2021xrx},
we note that the specific binary with DD2 \ac{eos} we used exhibits a larger deviation from the fitted 
relation for $\rhomax$, although it remains within the 90\% \ac{cl}. 

Combining the posteriors of the total mass and the tidal parameters with the results
above, we constrain the mass-radius diagram.
Figure~\ref{fig:eos_constraint} shows the $M(R)$ relation for the SLy ($\Mc=1.244295~\Msun$)
and BLh ($\Mc=1.227164~\Msun$) \acp{eos} (See Table~\ref{tab:IMPM}). The constrained areas are obtained with 
{\teobspa} for \ac{im}-only and {\teobnrpmw} for \ac{impm}. For \ac{pm}-only we use the {\teobnrpmw}~\ac{pe}
to compute $\Rmax$ and $\rhomax$, but we are not including the constraints on the mass and the tidal polarizability parameter.
We utilize the \acp{eos} prior sample of~\cite{Godzieba:2020tjn} and in our 90\% \ac{cl} 14372 \acp{eos}
are included. We additionally test scenarios
varying this amount, \ie~by doubling it, and obtain consistent results. 

Comparing the results for \ac{im}-only, \ac{pm}-only and \ac{impm} of Fig.~\ref{fig:eos_constraint} 
we can observe the impact of the constraints given by the {\teobnrpmw} waveform model alone and by
adding the quasiuniversal relations. The \ac{im}-only case is based only on the results from the 
\ac{gw} analysis, while the \ac{pm}-only is based solely on the inference on $f_2$ and the
quasiuniversal relations. The \ac{impm} case combines the two in a
stricter constraint. Therefore going from the case of the \ac{im}-only
(yellow) to the \ac{impm} (red) we additionally include the constraints
coming from the quasiuniversal relation to the \ac{gw} inference of $M$
and $\tilde{\Lambda}$. Looking at Fig.~\ref{fig:IM_IMPM}, we can see that the inference
of masses and tidal polarizability parameters is dominated by the \ac{im}. The case of the 
full-spectrum model {\teobnrpmw} does not
give additional constraints. Had we repeated the \ac{eos} sampling with the
\ac{impm} without any information from the quasiuniversal relations from
$f_2$, we would have obtained constraints identical to the \ac{im}-only.
The \ac{pm}, instead, gives information at higher densities and together with the \ac{im} produces 
a tighter constraint around the theoretical mass-radius curve, giving in particular a better estimation of
the maximum mass $\Mmax$ and the corresponding radius $\Rmax$. 
For \ac{impm}, we find the \acp{cl} to be $\Mmax = 2.03^{+0.07}_{-0.05}~\Msun$ and $\Rmax = 10.23 ^{+0.25}_{-0.12}{\rm km}$
for SLy and $\Mmax = 2.07^{+0.09}_{-0.08}~\Msun$ and $\Rmax = 10.75^{+0.32}_{-0.25}~{\rm km}$
for BLh.
The values we obtain are compatible with~\cite{Breschi:2021wzr} and the relative error 
is roughly the same. Compared with earlier analyses based on GW170817 
\cite{Breschi:2024qlc,Nathanail:2021tay,Koehn:2024set,Essick:2020flb}, an improvement by a factor three
in the uncertainties of the maximum mass is achieved with \ac{xg} detectors. 
For $\Rmax$, the precision enhances by a factor four~\cite{Rutherford:2024srk,Fan:2023spm}.
We note that in the region of $1.54\Msun\lesssim M \lesssim 1.87\Msun$, the constrained \ac{impm}
90\% confidence region is not totally consistent with the theoretical mass-radius relation for SLy and BLh. 
We attribute this slight incompatibility to the sample of \acp{eos}, that does not include \acp{eos}
exactly compatible with SLy or BLh at $M \lesssim 0.6~\Msun$.

Furthermore, Fig.~\ref{fig:lambda_m} reports the $\Lambda(M)$ curve derived from the combined \ac{impm}
analysis, evaluated for an equal-mass binary with BLh \ac{eos}, $M = 2.8\Msun$ and $\tilde{\Lambda} = 412$.
The resulting constraint on the tidal polarizability parameter as a function of the mass is 
particularly tight around the theoretical curve for BLh, highlighting both the accuracy of our method in recovering 
the theoretical prediction and its robustness in probing the dense matter properties of \acp{ns}.
From this analysis, we also extract constraints on the tidal polarizability 
parameter at $M=1.4\Msun$, finding that it lies within the range $289<\Lambda_{1.4}<529$.

Finally, Fig.~\ref{fig:p_rho} displays an example of direct inference of the pressure-density \ac{eos} relation.
The theoretical SLy curve lies within the 90\% \acp{cl} for both \ac{im} and \ac{impm} cases. 
We observe a similar behavior to that seen in the mass-radius curve: the confidence region is narrower in 
the region with $5\times 10^{14}~{\rm g~cm^{-3}} \lesssim \rho \lesssim 13\times 10^{14}~{\rm g~cm^{-3}}$, while it becomes broader 
at both lower and higher densities. 
As in the other inferences discussed, the \ac{impm} analysis improves the precision of the inference 
on $p(\rho)$ at higher densities.

\subsection{Prompt collapse}
\label{sec:pc}

\begin{figure}[t]
	\includegraphics[width=0.45\textwidth]{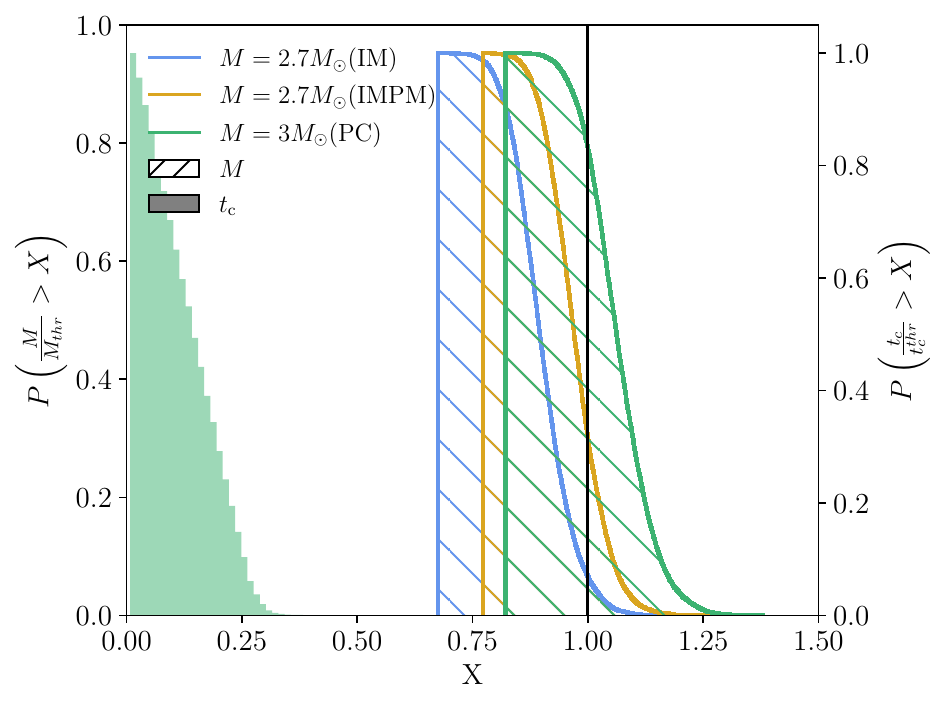}
	\caption{Cumulative posterior distributions of the ratio between the total mass and 
	the threshold mass for two different binaries. 
 	The probability of prompt collapse is represented by the fraction of posterior
	distribution that lies above the black line.
	In the case of SLy with $M=3~\Msun>M_{\rm thr}$ the probability of \ac{pc} is larger than
	80\%, in the other case ($M=2.7~\Msun<M_{\rm thr}$) the probability is smaller than 35\%.
	The difference between \ac{im} and \ac{impm} comes from the different ability to constrain the 
	maximum mass, between the two cases, as presented in Sec.~\ref{sec:highdens}.
	The colored area represents the probability distribution for the promptly collapsing binary
	of having a collapse time larger than the threshold collapse time, $t_{\rm c}^{\rm thr} = 2~{\rm ms}$. 
	The whole area lies below the black line, therefore the probability is vanishing.
	}
	\label{fig:thr}
\end{figure}

In this section we discuss the identification of \ac{pc}.
We initially analyze the lower part of Table~\ref{tab:IMPM},
for which we injected \ac{gw} signals with the \ac{im}-only model {\teobspa}, to simulate a \ac{pc}.
The recovery is performed both with \ac{im} and \ac{impm} templates. 
In these cases the \acp{bf} $\log\mathcal{B}_{\rm{IM}}^{\rm{IMPM}}$ 
are negative, with the smallest absolute value reaching $3.86^{+0.48}_{-0.48}$. 
Although these absolute values are smaller than the \acp{bf} in the upper part of the table, 
there is a strong evidence that the \ac{pc} hypothesis is favored, because the uncertainties are
low enough, implying the pipeline capability of reliably distinguishing \acp{pc}. 
We also note that these \acp{bf} are smaller, when comparing with the Bayesian \ac{pe} of 
Ref.~\cite{Breschi:2019srl}, performed with the time-domain full spectrum \teob{-NRPM}~model. 
Nevertheless, they remains sufficiently large to robustly confirm the occurrence of a \ac{pc}. 

While the \ac{bf} analysis demonstrates that our pipeline can detect \ac{pc} signatures 
under favorable conditions, its reliability diminishes as the signal weakens, particularly at 
larger distances. 
Notably, for \ac{xg} detectors, the \ac{snr} of the \ac{pm} typically reaches values on the order of tens, 
when the binary is located at a distance comparable to that of GW170817 event~\cite{LIGOScientific:2018mvr}, 
an exceptionally favorable case. However, the \ac{snr} drops
below the threshold ${\rm SNR_{thr}} \sim 7$ (see paper II) at larger distances. 
As a result, detecting and characterizing the \ac{pm} features becomes challenging, and even our pipeline 
may in principle not be able to reliably distinguish whether there is a \ac{pc} to a \ac{bh} 
or a \ac{ns} remnant is formed.
To address this degeneracy, we consider the alternative methods introduced in Sec.~\ref{sec:method:nrrel} to
infer \ac{pc}.

We present in Fig.~\ref{fig:thr} the probability of \ac{pc}
computed per~Eq.~\eqref{eq:prob_pc} for two equal-mass \acp{bns}, 
assuming SLy \ac{eos} ($M_{\rm thr} \simeq 2.8~\Msun$). The first binary has a total mass 
$M = 2.7~\Msun < M_{\rm thr}$ and the second has $M = 3.0~\Msun > M_{\rm thr}$.
In the former case, the probability of \ac{pc} 
is approximately $7\%$ for \ac{im}-only and $32\%$ for \ac{impm}.
This difference comes from the different inferences of the
maximum mass and the corresponding radius obtained with
\ac{im} and \ac{impm} analyses, see Fig.~\ref{fig:eos_constraint}. 
In the \ac{im} case $\Mmax$ is on average larger, resulting in a higher 
threshold mass. In contrast, the \ac{impm} analysis yields a more precise 
estimation of the maximum mass, leading to a smaller and more tightly constrained 
threshold mass. 

For the binary with $M = 3.0~\Msun$, the probability of \ac{pc} is $80\%$, strongly 
favoring the \ac{pc} scenario. 
Therefore, the first system is more likely to result in a long-lived \ac{ns} remnant, while the second
case supports the hypothesis of \ac{pc} to a \ac{bh}, as expected from the injection.

In addition, we estimate the values of the time of collapse, $t_{\rm c}$, with 
{\teobnrpmw} and we compute the probability of having $t_{\rm c}>t_{\rm c}^{\rm thr} = 2~{\rm ms}$.
For the first binary ($M = 2.7~\Msun$) we find $t_{\rm c} = 23^{+15}_{-14}~{\rm ms}$, while for 
the \ac{pc} injection ($M = 3.0~\Msun$), we find $t_{\rm c} = 0.28^{+0.26}_{-0.24}~{\rm ms}$.
From Fig.~\ref{fig:thr}, we notice that the probability $P\left(t_{\rm c}/t_{\rm c}^{\rm thr}>X\right)$
(green region) is vanishing for the latter binary.

Applying the same procedure to real observational data has the potential to place meaningful 
constraints on the properties of dense matter. Inferring the probability of \ac{pc} to \ac{bh} 
can provide valuable insights into the \ac{eos} characterization. This can help identify potential softening of
the \ac{eos}, which may arise from phase transitions at extreme densities or thermal 
effects~\cite{Radice:2016rys,Bauswein:2020xlt,Prakash:2023afe,Breschi:2023mdj,Fields:2023bhs}.

\section{Conclusion}
\label{sec:conclusion}

In this paper we presented a waveform model for the complete \ac{bns}
spectrum and its application to Bayesian parameter estimation with
next generation detectors. We focused on single signals in Einstein
Telescope's band and the inference of neutron star matter constraints.
Methodologically, our work introduces {\teobnrpmw}, a 
frequency domain waveform model that describes \acp{gw} from early inspiral up to kiloHertz
\ac{pm} and considers injection-recovery (mock) Bayesian
experiments to validate a pipeline that delivers inferences of various
extreme matter constraints.

As a preparatory step to full-spectrum analyses, we first considered
\ac{im}-only injections and study systematically the impact of the
choice of the initial frequency (down to $2~{\rm Hz}$), of the signal
\ac{snr} and the detector configuration on the inference.
Notably, we observe an improvement of approximately one order of magnitude in 
the precision of the chirp mass and mass ratio when lowering the initial frequency from 20 to 
2~Hz. The relative error on reduced tidal parameter improves by about a factor two. 
Varying the \ac{et} configuration from the triangular layout to the two-L 
configuration does not significantly impact the inference of the intrinsic parameters. 
Nevertheless, a clear improvement is seen in the measurement of the extrinsic parameters, 
particularly the declination, which is better constrained with the two-L configuration.
This improvement is primarily due to the geographical separation of the two detectors, 
one in Meuse-Rhine region and the other one in Sardinia, which enhance the ability to localize 
the source in the sky.

Next, we focused on the full spectrum analysis with the newly developed model {\teobnrpmw}.
Our pipeline effectively recovers both the injected inspiral segment and the peak 
frequency of the \ac{pm} signal. Other \ac{pm} features are also correctly retrieved, although their 
constraints are less stringent compared to those on the \ac{im} and
$f_2$. 
The posterior distributions of chirp mass, mass ratio and reduced tidal parameter, 
do not show a clear improvement when including the \ac{pm} spectrum. This is attributed to the 
\ac{pm} portion having an \ac{snr} approximately two orders of magnitude smaller than that of the 
\ac{im} section. Nevertheless, Bayesian model selection analysis favors the \ac{impm} model over the 
\ac{im}-only, already at postmerger SNR ${\sim}8$ (where the \ac{pm} detection threshold is 7).
This regimes proves to be essential for probing the \ac{ns} \acp{eos}, as we 
demonstrate that employing the full spectrum in the analyses leads to a substantial reduction 
of the credibility region on the mass-radius diagram, particularly in the high-density regime.

Using an \ac{eos} sampling with minimal hypotheses (causality, validity of general relativity), 
we showed that a single-event \ac{pm} signal plays a crucial role in constraining both
the maximum density and the radius corresponding to the maximum mass, through quasiuniversal 
relations that involves the peak frequency $f_2$. The inclusion of these posterior distributions
significantly reduces the 90\% credibility region of the mass-radius diagram, particularly at 
high densities. Therefore, the possibility to perform \acp{pe} of \ac{gw} signals detected with 
\ac{et}, that also contains \ac{pm} features, enables more precise constraints on the maximum 
mass and the corresponding radius. In our study, we demonstrate relative errors 
smaller than 8\% and 5\% for the maximum mass and the relative neutron
star radius, respectively.
A novel result presented in this paper is the tight constraint on the $\Lambda(M)$ curve, where 
the 90\% \ac{cl} closely follows the theoretical model. 
Similarly, the inference of the pressure-density relation, especially that obtained through the 
\ac{impm} analysis, yields significantly tight constraints at high densities, again closely tracking 
the theoretical prediction. 
This approach enhances the precision in extracting information about the properties of nuclear 
matter under extreme conditions. 

Finally, we analyze the inference of prompt collapse scenario.
Bayesian model selection with \ac{im}-only and \ac{impm} recoveries
can identify \ac{pc} for \ac{pm} detections at or above the SNR threshold.
However, other complementary approaches can be utilized.
We extended the approach of~\citet{Agathos:2019sah} to infer the 
probability of \ac{pc}. Our analysis highlights that it is
possible to robustly determine whether a BNS undergoes \ac{pc} or forms a \ac{ns} 
remnant that emits \ac{pm} signals, 
even if the latter is not detectable due to a weak \ac{pm} \ac{snr}.

Our work has important limitations that should be considered for future work. 
First, in both \ac{im}-only and full spectrum analyses, we injected nonspinning binary systems. 
In the future, it would be valuable to access the ability of our pipeline to accurately 
recover spin components, despite the degeneracy between spin, mass ratio and tidal 
polarizability parameters. 
An ongoing study is already investigating \ac{pe} without fixing any parameters in the 
\ac{impm} scenario. This includes the inference of the spin components as well as 
additional extrinsic parameters of the binary, beyond just the luminosity distance. 
Moreover, our current study focuses only on the dominant (2,2) mode. 
Including also higher-order modes would help, for example to address the degeneracy between 
luminosity distance and the inclination. {\teobspa} already incorporates higher modes,
whereas their implementation in \model~is still undergoing. 
Given that our injections are based on \ac{gw} signals from \ac{xg} of detectors, we must 
also account for the long duration of such signal. In this regime, Earth's rotation is not 
negligible, and the increased sensitivity of the detector implies that multiple 
signals may overlap in the data stream. Consequently, future analysis techniques must
address challenges in signal disentanglement and noise mitigation~\cite{Abac:2025saz,Cireddu:2023ssf,Liu:2024jna,Wong:2024hes}. 
Lastly, in our analyses, we used the same model for both injection and recovery. 
However, \citet{Gamba:2020wgg} showed that for signals with \ac{snr} $>80$, 
using different waveform models for injection and recovery introduces significant systematic
errors. This highlights the importance of waveform systematics studies and motivates the 
development of more robust cross-model validation strategies.

\section*{Acknowledgments}

The authors thank Nishkal Rao for technical contributions and discussion.
G.H. and S.B. acknowledge support by the EU Horizon under ERC
Consolidator Grant, No. InspiReM-101043372. 
S.B. and M.B. acknowledge 
support by the EU H2020 under ERC Starting
Grant, no.~BinGraSp-714626.
S.B. acknowledges support from the Deutsche Forschungsgemeinschaft (DFG)
project MEMI (BE 6301/2-1 Projektnummer: 443239082) and GROOVHY (BE 6301/5-1 Projektnummer: 523180871).
M.B. and R.G. acknowledges support from the Deutsche Forschungsgemeinschaft
(DFG) under Grant No. 406116891 within the Research Training Group
RTG 2522/1.
R.G. acknowledges support from NSF Grant No. PHY-2020275 [Network for Neutrinos, Nuclear Astrophysics, 
and Symmetries (N3AS)].

Computations were performed on SuperMUC-NG at the Leibniz-Rechenzentrum
(LRZ) Munich and and on the national HPE Apollo Hawk at the High
Performance Computing Center Stuttgart (HLRS). The authors acknowledge
the Gauss Centre for Supercomputing e.V. (\url{www.gauss-centre.eu})
for funding this project by providing computing time on the GCS
Supercomputer SuperMUC-NG at LRZ (allocations {\tt pn76li}, {\tt
  pn36jo} and {\tt pn68wi}). The authors acknowledge HLRS for funding
this project by providing access to the supercomputer HPE Apollo Hawk
under the grant number INTRHYGUE/44215 and MAGNETIST/44288. 
Postprocessing and development runs were performed on the ARA cluster
at Friedrich Schiller University Jena. The ARA cluster is funded in
part by DFG grants No. INST 275/334-1 FUGG and No. INST 275/363-1 FUGG, and
ERC Starting Grant, grant agreement no. BinGraSp-714626.

\section{Data availability}

The waveform model developed in this work, {\teobnrpmw}, 
is implemented in \bajes~and the software is publicly available at:

\url{https://github.com/matteobreschi/bajes} 

\noindent \teob{} is publicly available at:

\url{https://bitbucket.org/teobresums/teobresums}.

%%______________________________________________________________


\appendix

\section{Full spectrum model}
\label{app:wvf}

\begin{table}[t]
	\centering    
	\caption{Summary of the \ac{nr} simulations used to validate {\teobnrpmw}~template.
		The first column reports the nature of the remnant.
		The following seven columns show the \ac{nr} data properties,
		\ie~ \ac{eos}, reference, total mass $M$, mass ratio $q$, tidal polarizability $\kt$,
		\ac{pm} peak frequency $f_2$ and the time of \ac{bh} collapse $t_{\rm c}$.}
		\begin{tabular}{c|ccccccc}
			\hline
			\hline
            Morph. & EOS & Ref.& $M$ & $q$ & $\kt$& $f_2$& $t_{\rm c}$ \\
            & & & $[\Msun]$ & & &$[{\rm kHz}]$&$[{\rm ms}]$\\
            \hline
            \hline
			\multirow{2}{*}{Short-lived} & 	
            SFHo & \cite{Bernuzzi:2015opx} & $2.7$ & $1.0$ & $79$ & $3.42$ & $8.0$ \\
			& BLQ & \cite{Prakash:2021wpz} & 2.8 & 1.0 & 81 & 2.68 & 2.7 \\
			\hline
			\multirow{2}{*}{Long-lived} & 	
            SLy & \cite{Breschi:2019srl} & $2.6$ & $1.0$ & $92$ & $3.13$ & $21$ \\
			& LS220 & \cite{Perego:2019adq} & 2.6 & 1.167 & 168 & 2.80 & 35 \\
			\hline
			\multirow{2}{*}{Tidally disrupted} & 	
            BLh & \cite{Bernuzzi:2020txg} & $2.84$ & $1.664$ & $97$ & $3.27$ & $18$ \\
			& DD2 & \cite{Bernuzzi:2024mfx} & 2.88 & 1.666 & 131 & 2.58 & 34 \\
			\hline
			\hline
	\end{tabular}
	\label{tab:inject}
\end{table}

\begin{figure}[t]
	\centering 
	\includegraphics[width=0.49\textwidth]{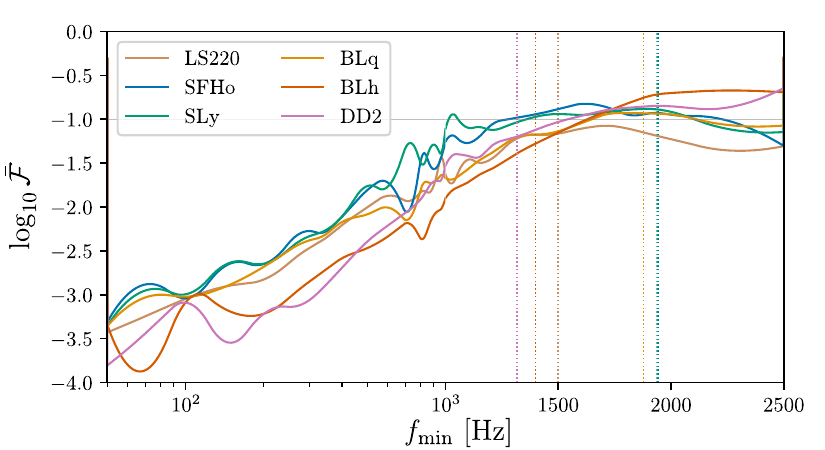}
	\caption{Unfaithfulness between hybrid NR-EOB waveforms 
		and {\teobnrpmw} model as function of the 
		lower cutoff frequency $f_{\rm min}$.
		The values are computed employing ET-D~\cite{Hild:2010id,Hild:2011np}
		sensitivity curve and setting the
		upper cutoff frequency to $8~{\rm kHz}$.
		The vertical dotted lines refer to the value
		of the instantaneous \ac{nr} frequency at merger
		for each case.
	}
	\label{fig:impm_fbar}
\end{figure}

\begin{figure}[t]
	\centering 
	\includegraphics[width=0.49\textwidth]{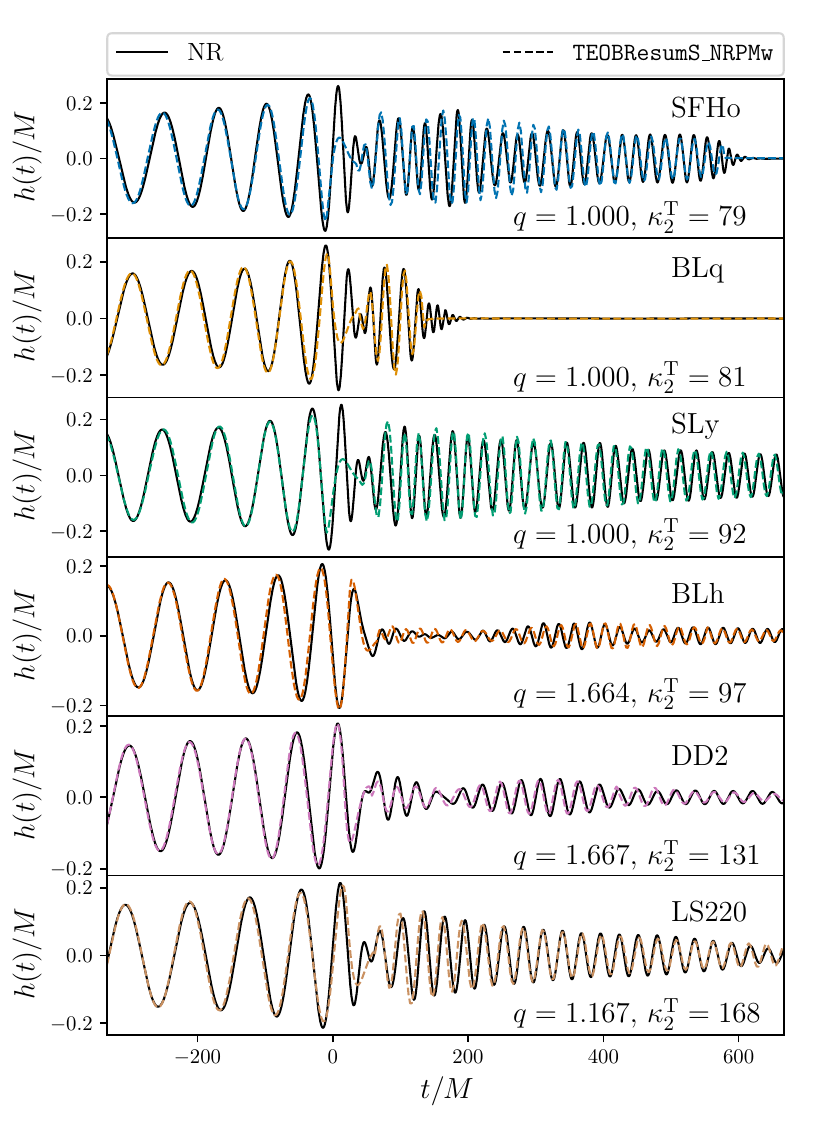}
	\caption{Comparison between \ac{nr} data (black solid lines) and 
		{\teobnrpmw} model (colored dashed lines).
		The reported \ac{nr} templates correspond to the 
		validation set listed in Tab.~\ref{tab:inject}.	
		The time-domain {\teobnrpmw} waveforms
		are estimated employing inverse \ac{fft}.
		For each \ac{nr} simulation, 
		we show the corresponding \ac{eos}, the mass ratio $q$
		and the tidal polarizability $\kt$.
		The waveforms are aligned on the 
		segment $t/M\le -200$.
	}
	\label{fig:impm_wf}
\end{figure}

We provide here some details on the construction and the validation of the complete waveform 
model {\teobnrpmw}~employed in our work.

The complete model is obtained by attaching the \ac{eob} {\teobspa} to \model~at the merger.  
Within this framework, the parameters $\{\phi_{\rm PM},t_0\}$
determine the initial phase and time of the \ac{pm} segment (see paper I) 
and are set to the phase and time of merger from the {\teobspa}~waveform, $\phi_{\rm mrg}$
and $t_{\rm mrg}$.
In order to perform a cleaner attachment and avoid double-counting,
when we employ {\model} for the complete \ac{impm} model,
the fusion wavelet is set to zero, \ie~$\wavelet_{\rm fus}=0$,
since the \ac{eob} model is expected to include the merger portion of data.

The validation of the complete model has been performed using 6 exemplary \ac{nr} simulations 
from the \core~database~\cite{Gonzalez:2022mgo} computed with the {\scshape thc} code~\cite{Bernuzzi:2015opx,Perego:2019adq,Bernuzzi:2020txg,Prakash:2021wpz,Kashyap:2021wzs} 
and listed in Tab.~\ref{tab:inject}. 
Note this set of simulations 
is not included in the calibration set.
The data are chosen to cover a sufficiently large variety of phenomenologies; 
in particular, the validation set includes four long-lived cases, 
two of which with $q>1.5$ showing tidal disruption,
and two short-lived cases, 
one of which includes deconfined quark matter.
Moreover, \ac{nr} data are hybridized with the time-domain \teob{} model, in order to 
extended the waveforms in the early inspiral regime,
not included in typical \ac{nr} simulations.
Also for these studies, we employ ET-D sensitivity curve~\cite{Hild:2010id,Hild:2011np}
and we minimize over the additional set of parameters 
$\{\freeparams,\recalib_{\rm PM}\}$ for {\model}.
Employing only the \ac{pm} portion of data and {\model},
we obtain $\log_{10}\Fbar$ between 
$-1.35$ and $-0.95$
over all the validation set of 6 binaries.
In particular, short-lived cases (SFHo and BLQ) 
reach $\log_{10}\Fbar \lesssim-1.25$;
while, large mass ratio binaries (BLh and DD2) 
recover worst matches, \ie~ $\log_{10}\Fbar \simeq-0.95$.

Figure~\ref{fig:impm_fbar} shows the 
unfaithfulness computed between {\teobnrpmw} model
and the hybrid NR-EOB waveforms,
where the inner products are evaluated 
over the frequency range $[f_{\rm min}, 8~{\rm kHz}]$.
For $f_{\rm min} < 100~{\rm Hz}$,
the model have good agreement 
with \ac{nr} data as shown by $\log_{10}\Fbar\lesssim -3$,
comparable to Ref.~\cite{Gamba:2020ljo}.
Approaching $f_{\rm mrg}$,
the recovered $\Fbar$ shows an increasing trend,
with peaks above $10^{-1}$.
Subsequently, for $f_{\rm min} > 1.5~{\rm kHz}$,
the recovered $\Fbar$ are comparable with the values 
obtained only with the \ac{pm} portion of data, \ie~ $\log_{10}\Fbar\lesssim -1$.
In order to further inspect this behavior,
we show in Fig.~\ref{fig:impm_wf} the comparison 
between the time-domain \ac{nr} data and the corresponding 
{\teobnrpmw} templates, making use of the inverse \ac{fft}.
The inspiral \ac{eob} template matches \ac{nr} data accurately up to merger, \ie~ $t=0$;
however, 
considerable phase differences arise after this moment in the two models.
These errors can be related to the \ac{spa} 
and to the absence of next-to-quasicircular (NQC) corrections~\cite{Riemenschneider:2021ppj,Albertini:2021tbt}
in the frequency-domain \ac{eob} template. These corrections appear to be 
more relevant for equal-mass binaries and they 
motivate additional studies for the development
and the calibration of accurate EOB-NR waveform models.

\section{Mock Bayesian inferences}
\label{app:inf}

In this section, we summarize the basic inference techniques implemented in
\bajes~pipeline~\cite{Breschi:2021wzr}. The main novelty introduced for this work is 
the implementation of the relative binning technique from~\cite{Dai:2018dca, Zackay:2018qdy}.

\subsection{Bayesian framework}
\label{app:inf:bajes}

Bayesian analysis is a primary technique to perform parameter estimation. It based 
on the evaluation of the posterior distribution $p(\boldsymbol{\theta}|\boldsymbol{d},H)$, 
through Bayes' theorem,
\be
	p(\params|\data,H_{\rm S}) = \frac{p(\data|\params,H_{\rm S}) p(\params|H_{\rm S})}{p(\data|H_{\rm S})},
    \label{eq:bayes}
\ee
where the prior distribution $p(\params|H_{\rm S})$ represents our knowledge about the parameters
we estimate, $p(\data|H_{\rm S})$ is the evidence and $p(\data|\params,H_{\rm S})$ is
called likelihood function. 
For a fixed set of data $\data(t) $, the likelihood function is given by
\be
	\label{eq:likelihood}
	\log p(\data|\params,H_{\rm S}) \propto -\half \sum_i \big(d-h(\params)\big|d-h(\params)\big)_i\,,
\ee
where $H_{\rm S}$ represents the signal hypothesis,
the subscript $i$ runs over the employed detectors 
(\ie~ it denotes that the inner products are estimated 
with the corresponding data series, projected waveform 
and \ac{psd} for each detector),
and the parameter vector $\params $ corresponds to the combination 
$\params=\{ \params_{\rm bin},\freeparams, \recalib_{\rm fit}, \extparams\}$
of 
the inspiral parameters $\params_{\rm bin}$,
the unconstrained \ac{pm} parameters $\freeparams$,
the recalibration coefficients $\recalib_{\rm fit}$
and the extrinsic \ac{gw} parameters $\extparams=\{D_L, \iota , \alpha, \delta, \psi, t_{\rm mrg}, \phi_{\rm mrg}\}$.
The inner product in Eq.~\eqref{eq:likelihood} is 
\be
	(a|b) = \int_{f_0}^{f_{\rm max}} \dfrac{a^*(f)b(f)}{S_n(f)}df,
	\label{eq:innerprod}
\ee
with $S_n(f)$ the \ac{psd} of the noise segment.

Resorting to nested sampling algorithm~\cite{Skilling:2006gxv},
it is possible to accurately
estimate the evidence integral, \ie
\be
\label{eq:evidence}
p(\data|H_{\rm S})= \int p(\data|\params,H_{\rm S}) \,p(\params|H_{\rm S}) \,\d\params \,,
\ee
that allows us to perform model selection between two model 
hypotheses A and B, computing the \ac{bf} 
of the signal hypothesis A against the signal hypothesis B as
\be
	\mathcal{B}_{\rm{A}}^{\rm{B}} = \frac{p(d|H_{\rm B})}{p(d|H_{\rm A})}.
\ee
When $\mathcal{B}_{\rm{A}}^{\rm{B}} > 1$ ($<1$), the model hypothesis B~(A) is favored.
Furthermore, we introduce the \ac{snr}
from Eq.~\eqref{eq:innerprod} as
\be
\label{eq:snr}
\rho(h) = \frac{(d_i|h)}{\sqrt{(h|h)}}\,,
\ee
where the definition can be extended to multiple detectors 
employing quadrature summation.

For a more detailed discussion on the Bayesian framework employed in \ac{gw} data analysis
we remand to~\cite{Veitch:2014wba,LIGOScientific:2019hgc,
Breschi:2021wzr}.

\subsection{Relative binning}
\label{app:inf:rb}

Relative binning~\cite{Dai:2018dca,Zackay:2018qdy,Krishna:2023bug} is an acceleration method for \ac{pe}. 
The most computationally expensive part of a \ac{pe} is the evaluation of the likelihood
function~\eqref{eq:likelihood} with waveforms $h(f,\params)$ in the parameter space, that are 
small perturbations of the best-fit waveform.
Neighboring waveforms in the frequency domain are very similar to each other, 
therefore one can work directly with the ratio with respect to a fiducial waveform $h_0(f)$,
\be
\label{eq:ratioRB}
	r(f) = \dfrac{h(f)}{h_0(f)}.
\ee 
Since this ratio is smooth, it can be approximated by a piecewise
linear function~\cite{Zackay:2018qdy}
\be
\label{eq:ratioRB_lin}
	r(f) = r_0(h,b) + r_1(h,b)(f-f_m(b)), 
\ee 
where $b$ represents a frequency bin and $f_m$ is the mean frequency of the bin. 
Some components of the inner products involved in the likelihood computation
can be precomputed to improve the efficiency
\be
\label{eq:sumdata}
\begin{split}
	A_0 &= \frac{4}{T} \sum_{f\in b}\dfrac{d(f)h_0^*(f)}{S_n}, \\
	A_1 &= \frac{4}{T} \sum_{f\in b}\dfrac{d(f)h_0^*(f)}{S_n}(f - f_m(b)),\\
	B_0 &= \frac{4}{T} \sum_{f\in b}\dfrac{|h_0(f)|^2}{S_n}, \\
	B_1 &= \frac{4}{T} \sum_{f\in b}\dfrac{|h_0(f)|^2}{S_n}(f - f_m(b))\,.
\end{split}
\ee
Thus, during the \ac{pe} it is sufficient to evaluate $r_0(h,b)$ and $r_1(h,b)$ at the bin edges
and compute the inner products for the likelihood as
\be
\label{eq:dh-hhRB}
\begin{split}
	& (d|h) = \sum_b (A_0(b)r_0^*(h,b) + A_1(b)r_1^*(h,b)), \\
	& (h|h) = \sum_b (B_0(b)|r_0(h,b)|^2 + 2B_1(b)\Re[r_0(h,b)r_1^*(h,b)]).
\end{split}
\ee
We then employ the binning criterion of~\cite{Zackay:2018qdy} with $\epsilon = 0.01$. \\

Relative binning technique is effective for waveform 
analyses up to the merger part, as the frequency domain representations are smooth in 
this regime. Instead, in the \ac{pm} regime, the waveform structure becomes highly 
complex and rapidly varying in frequency, rendering relative binning inefficient and insufficiently 
accurate. Nevertheless, \model~is fast, being phenomenological, therefore
acceleration techniques are not critical as in the case of {\teobspa}. 
In our full spectrum analyses, we thus employ relative binning only in the \ac{im} portion 
and we perform standard \ac{pe} for the \ac{pm} part of the spectrum. 
To implement this approach, we introduce cutoff frequency $f_{\rm cut} = 2048~{\rm Hz}$.
The likelihood is computed separately for the two regimes: using relative binning for 
frequencies below $f_{\rm cut}$ (\ac{im}) and using the full, unaccelerated Bayesian \ac{pe} for frequencies
above the cutoff (\ac{pm}). The total likelihood is obtained by summing the two contributions,
cf. Eq~\eqref{eq:likel_sum}.
This hybrid approach enables efficient and accurate \ac{pe} for \ac{gw} signals spanning 
both the \ac{im} and \ac{pm} regimes.\\

The precomputation of some terms and the evaluation of the waveforms only on the bin edges 
enable \ac{pe} of coalescing compact binaries in a very short amount of time: it is possible to run
a \ac{pe} on a single CPU in 44 minutes for a binary black holes, using the 
\teob{}~template, and in 275 minutes for a \ac{bns} with \texttt{TaylorF2}. 
In the case of full spectrum \acp{bns} waveforms with {\teobnrpmw}, we run on a cluster with 192
CPUs in about 40 hours, making use of the MPI-parallelization of \bajes~\cite{Breschi:2021wzr}. \\

\section{Results}
\label{app:data}
 
We report in Table~\ref{tab:IM2} the results of the injection recovery with 
\ac{im}-only {\teobspa}~template relative to Fig.~\ref{fig:lambda_SNR_dl} and~\ref{fig:lambda_SNR_f}.

\onecolumngrid
\vspace*{20pt}
\LTcapwidth=\linewidth
\begin{longtable}{cccccccc|cccccc}
	\caption{Summary of the \acp{pe} with {\teobspa} for \acp{bns}.
	The first eight columns report the injected parameters and the last six columns report the recovered values, with the
	median of the posterior distributions and the $90\%$ credibility regions.} \\
		\hline
		\hline
		\multicolumn{8}{c|}{Injected properties}&\multicolumn{6}{c}{Recovered values}\\
		\hline
		$\Mc~[\Msun]$ & $q$ & $\tilde{\Lambda}$ & $\delta\tilde{\Lambda}$ & $\chi_{\rm eff}$ & $D_L~[{\rm Mpc}]$ & $f_{0}$ & SNR & $\Mc~[\Msun]$ & $q$ & $\tilde{\Lambda}$ & $\delta\tilde{\Lambda}$ & $\chi_{\rm eff}$ & $D_L~[{\rm Mpc}]$\\
		\hline 		\endfirsthead
		&  &  &  &  &  & 5 & 2452 & $1.3447868^{+0.0000005}_{-0.0000005}$ & $1.212^{+0.005}_{-0.005}$ & $172^{+4}_{-6}$ & $40^{+30}_{-40}$ & $-0.0000^{+0.0003}_{-0.0004}$ & $38.6^{+0.8}_{-0.8}$ \\
		&  &  &  &  & 40 & 10 & 1775 & $1.344787^{+0.000003}_{-0.000002}$ & $1.21^{+0.02}_{-0.01}$ & $171^{+4}_{-7}$ & $40^{+30}_{-30}$ & $0.0000^{+0.0009}_{-0.0007}$ & $38.2^{+1.0}_{-0.8}$ \\
		&  &  &  &  &  & 20 & 1431 & $1.344788^{+0.000012}_{-0.000011}$ & $1.22^{+0.05}_{-0.05}$ & $170^{+6}_{-12}$ & $30^{+40}_{-40}$ & $0.000^{+0.003}_{-0.003}$ & $38.2^{+1.2}_{-1.1}$ \\
		&  &  &  &  &  & 5 & 1226 & $1.3447868^{+0.0000010}_{-0.0000009}$ & $1.213^{+0.010}_{-0.009}$ & $170^{+6}_{-9}$ & $40^{+30}_{-50}$ & $0.0001^{+0.0005}_{-0.0004}$ & $76^{+2}_{-2}$ \\
		1.3447868 & 1.212 & 172 & 30 & 0 & 80 & 10 & 887 & $1.344788^{+0.000005}_{-0.000004}$ & $1.22^{+0.03}_{-0.03}$ & $168^{+7}_{-11}$ & $30^{+40}_{-50}$ & $0.000^{+0.002}_{-0.002}$ & $75^{+3}_{-3}$ \\
		&  &  &  &  &  & 20 & 716 & $1.34479^{+0.00002}_{-0.00002}$ & $1.22^{+0.08}_{-0.08}$ & $168^{+7}_{-16}$ & $30^{+40}_{-50}$ & $0.000^{+0.005}_{-0.004}$ & $75^{+3}_{-3}$ \\
		&  &  &  &  &  & 5 & 817 & $1.344787^{+0.0000014}_{-0.0000015}$ & $1.216^{+0.014}_{-0.015}$ & $167^{+8}_{-13}$ & $30^{+40}_{-50}$ & $0.0001^{+0.0008}_{-0.0007}$ & $113^{+4}_{-4}$ \\
		&  &  &  &  & 120 & 10 & 592 & $1.344788^{+0.000007}_{-0.000006}$ & $1.22^{+0.04}_{-0.04}$ & $166^{+9}_{-17}$ & $30^{+40}_{-60}$ & $0.000^{+0.002}_{-0.002}$ & $113^{+5}_{-5}$ \\
		&  &  &  &  &  & 20 & 477 & $1.34478^{+0.00004}_{-0.00002}$ & $1.19^{+0.14}_{-0.13}$ & $168^{+7}_{-18}$ & $30^{+40}_{-60}$ & $-0.002^{+0.008}_{-0.005}$ & $112^{+6}_{-6}$ \\
		\hline 		&  &  &  &  &  & 5 & 2391 & $1.3040200^{+0.0000005}_{-0.0000005}$ & $1.222^{+0.006}_{-0.005}$ & $193^{+4}_{-7}$ & $50^{+30}_{-40}$ & $-0.0000^{+0.0003}_{-0.0003}$ & $38.6^{+0.9}_{-0.9}$ \\
		&  &  &  &  & 40 & 10 & 1731 & $1.304020^{+0.000002}_{-0.000002}$ & $1.22^{+0.02}_{-0.02}$ & $192^{+5}_{-8}$ & $50^{+30}_{-40}$ & $0.0001^{+0.0009}_{-0.0008}$ & $38.2^{+1.0}_{-0.9}$ \\
		&  &  &  &  &  & 20 & 1396 & $1.304023^{+0.000009}_{-0.000011}$ & $1.24^{+0.04}_{-0.05}$ & $190^{+7}_{-11}$ & $40^{+40}_{-40}$ & $0.001^{+0.003}_{-0.003}$ & $38.2^{+1.0}_{-1.0}$ \\
		&  &  &  &  &  & 5 & 1195 & $1.3040201^{+0.0000009}_{-0.0000009}$ & $1.224^{+0.009}_{-0.009}$ & $190^{+6}_{-10}$ & $50^{+30}_{-50}$ & $0.0001^{+0.0005}_{-0.0005}$ & $76^{+2}_{-2}$ \\
		1.3040200 & 1.222 & 192 & 35 & 0 & 80 & 10 & 865 & $1.304021^{+0.000004}_{-0.000005}$ & $1.23^{+0.03}_{-0.03}$ & $189^{+8}_{-14}$ & $40^{+40}_{-60}$ & $0.000^{+0.002}_{-0.002}$ & $76^{+3}_{-3}$ \\
		&  &  &  &  &  & 20 & 698 & $1.30402^{+0.00002}_{-0.00002}$ & $1.22^{+0.09}_{-0.09}$ & $189^{+7}_{-19}$ & $40^{+40}_{-60}$ & $-0.000^{+0.006}_{-0.004}$ & $75^{+3}_{-3}$ \\
		&  &  &  &  &  & 5 & 797 & $1.3040203^{+0.0000014}_{-0.0000013}$ & $1.225^{+0.014}_{-0.013}$ & $187^{+9}_{-15}$ & $40^{+40}_{-60}$ & $0.0002^{+0.0007}_{-0.0006}$ & $114^{+5}_{-4}$ \\
		&  &  &  &  & 120 & 10 & 577 & $1.304021^{+0.000006}_{-0.000005}$ & $1.23^{+0.04}_{-0.04}$ & $186^{+9}_{-14}$ & $30^{+40}_{-50}$ & $0.000^{+0.002}_{-0.002}$ & $114^{+5}_{-6}$ \\
		&  &  &  &  &  & 20 & 465 & $1.30402^{+0.00003}_{-0.00002}$ & $1.23^{+0.13}_{-0.14}$ & $186^{+10}_{-21}$ & $30^{+50}_{-60}$ & $0.000^{+0.008}_{-0.006}$ & $113^{+6}_{-6}$ \\
		\hline 		&  &  &  &  &  & 5 & 2344 & $1.2742306^{+0.0000004}_{-0.0000004}$ & $1.500^{+0.003}_{-0.003}$ & $233^{+6}_{-6}$ & $50^{+20}_{-20}$ & $-0.0001^{+0.0003}_{-0.0005}$ & $38.5^{+0.9}_{-0.7}$ \\
		&  &  &  &  & 40 & 10 & 1697 & $1.274232^{+0.000002}_{-0.000002}$ & $1.505^{+0.011}_{-0.011}$ & $230^{+9}_{-10}$ & $70^{+30}_{-30}$ & $0.0001^{+0.0009}_{-0.0009}$ & $38.1^{+1.2}_{-0.9}$ \\
		&  &  &  &  &  & 20 & 1369 & $1.274231^{+0.000008}_{-0.000014}$ & $1.50^{+0.02}_{-0.04}$ & $232^{+10}_{-12}$ & $80^{+20}_{-30}$ & $0.000^{+0.002}_{-0.004}$ & $38.1^{+1.3}_{-1.1}$ \\
		&  &  &  &  &  & 5 & 1172 & $1.2742306^{+0.0000009}_{-0.0000010}$ & $1.501^{+0.006}_{-0.007}$ & $231^{+9}_{-11}$ & $80^{+30}_{-40}$ & $0.0001^{+0.0006}_{-0.0005}$ & $76^{+2}_{-2}$ \\
		1.2742305 & 1.5 & 232 & 75 & 0 & 80 & 10 & 849 & $1.274231^{+0.000005}_{-0.000005}$ & $1.50^{+0.02}_{-0.02}$ & $230^{+10}_{-20}$ & $80^{+30}_{-40}$ & $0.001^{+0.002}_{-0.002}$ & $75^{+3}_{-3}$ \\
		&  &  &  &  &  & 20 & 684 & $1.274230^{+0.000014}_{-0.000017}$ & $1.50^{+0.04}_{-0.05}$ & $230^{+10}_{-20}$ & $80^{+30}_{-40}$ & $0.000^{+0.004}_{-0.004}$ & $75^{+3}_{-3}$ \\
		&  &  &  &  &  & 5 & 782 & $1.2742308^{+0.0000013}_{-0.0000013}$ & $1.502^{+0.009}_{-0.008}$ & $230^{+10}_{-20}$ & $80^{+30}_{-40}$ & $0.0001^{+0.0007}_{-0.0007}$ & $114^{+5}_{-5}$ \\
		&  &  &  &  & 120 & 10 & 566 & $1.274232^{+0.000007}_{-0.000006}$ & $1.51^{+0.03}_{-0.03}$ & $220^{+10}_{-20}$ & $70^{+30}_{-50}$ & $0.000^{+0.003}_{-0.002}$ & $112^{+5}_{-5}$ \\
		&  &  &  &  &  & 20 & 456 & $1.27423^{+0.00003}_{-0.00002}$ & $1.49^{+0.08}_{-0.06}$ & $230^{+20}_{-30}$ & $70^{+30}_{-50}$ & $-0.001^{+0.007}_{-0.005}$ & $111^{+6}_{-5}$ \\
		\hline 		&  &  &  &  &  & 5 & 2238 & $1.2037856^{+0.0000004}_{-0.0000004}$ & $1.333^{+0.004}_{-0.004}$ & $335^{+7}_{-9}$ & $90^{+40}_{-40}$ & $0.0000^{+0.0004}_{-0.0003}$ & $38.5^{+1.1}_{-0.8}$ \\
		&  &  &  &  & 40 & 10 & 1620 & $1.203786^{+0.000002}_{-0.000002}$ & $1.334^{+0.014}_{-0.014}$ & $336^{+10}_{-13}$ & $90^{+50}_{-50}$ & $-0.0001^{+0.0010}_{-0.0010}$ & $38.3^{+1.1}_{-1.0}$ \\
		&  &  &  &  &  & 20 & 1307 & $1.203783^{+0.000011}_{-0.000008}$ & $1.32^{+0.05}_{-0.03}$ & $340^{+10}_{-20}$ & $90^{+50}_{-50}$ & $-0.001^{+0.003}_{-0.002}$ & $37.9^{+1.3}_{-1.2}$ \\
		&  &  &  &  &  & 5 & 1119 & $1.2037856^{+0.0000008}_{-0.0000007}$ & $1.333^{+0.007}_{-0.006}$ & $335^{+10}_{-14}$ & $100^{+40}_{-60}$ & $-0.0001^{+0.0006}_{-0.0005}$ & $76^{+3}_{-2}$ \\
		1.2037856 & 1.333 & 334 & 75 & 0 & 80 & 10 & 810 & $1.203786^{+0.000004}_{-0.000005}$ & $1.34^{+0.03}_{-0.03}$ & $330^{+20}_{-20}$ & $80^{+60}_{-80}$ & $0.001^{+0.002}_{-0.00}$ & $76^{+3}_{-3}$ \\
		&  &  &  &  &  & 20 & 654 & $1.20378^{+0.00002}_{-0.00002}$ & $1.33^{+0.09}_{-0.07}$ & $330^{+10}_{-30}$ & $90^{+50}_{-70}$ & $-0.000^{+0.006}_{-0.004}$ & $75^{+3}_{-3}$ \\
		&  &  &  &  &  & 5 & 746 & $1.2037857^{+0.0000012}_{-0.0000011}$ & $1.334^{+0.011}_{-0.009}$ & $330^{+10}_{-20}$ & $110^{+40}_{-70}$ & $0.0001^{+0.0007}_{-0.0007}$ & $114^{+5}_{-5}$ \\
		&  &  &  &  & 120 & 10 & 540 & $1.203787^{+0.000006}_{-0.000006}$ & $1.34^{+0.03}_{-0.04}$ & $330^{+20}_{-30}$ & $80^{+70}_{-80}$ & $0.001^{+0.002}_{-0.002}$ & $111^{+6}_{-5}$ \\
		&  &  &  &  &  & 20 & 436 & $1.20378^{+0.00003}_{-0.00002}$ & $1.33^{+0.13}_{-0.09}$ & $330^{+20}_{-40}$ & $80^{+60}_{-90}$ & $-0.001^{+0.010}_{-0.006}$ & $112^{+6}_{-6}$ \\
 		\hline 		\pagebreak
		\hline
		&  &  &  &  &  & 5 & 2293 & $1.2395655^{+0.0000005}_{-0.0000004}$ & $1.117^{+0.009}_{-0.008}$ & $385^{+4}_{-7}$ & $80^{+70}_{-90}$ & $0.0000^{+0.0002}_{-0.0002}$ & $38.5^{+0.9}_{-0.8}$ \\
		&  &  &  &  & 40 & 10 & 1660 & $1.239566^{+0.000002}_{-0.000002}$ & $1.12^{+0.02}_{-0.02}$ & $385^{+4}_{-8}$ & $80^{+80}_{-90}$ & $0.0001^{+0.0007}_{-0.0007}$ & $38.3^{+1.0}_{-0.9}$ \\
		&  &  &  &  &  & 20 & 1339 & $1.239564^{+0.000012}_{-0.000005}$ & $1.10^{+0.08}_{-0.05}$ & $385^{+4}_{-13}$ & $60^{+80}_{-100}$ & $-0.001^{+0.003}_{-0.001}$ & $38.1^{+1.2}_{-1.0}$ \\
		&  &  &  &  &  & 5 & 1146 & $1.2395657^{+0.0000008}_{-0.0000008}$ & $1.121^{+0.015}_{-0.015}$ & $381^{+7}_{-13}$ & $50^{+90}_{-110}$ & $0.0001^{+0.0005}_{-0.0004}$ & $76^{+3}_{-2}$ \\
		1.2395655 & 1.117 & 385 & 35 & 0 & 80 & 10 & 830 & $1.239566^{+0.000004}_{-0.000003}$ & $1.12^{+0.05}_{-0.05}$ & $382^{+7}_{-15}$ & $50^{+90}_{-90}$ & $0.000^{+0.002}_{-0.001}$ & $76^{+3}_{-3}$ \\
		&  &  &  &  &  & 20 & 670 & $1.239566^{+0.000018}_{-0.000010}$ & $1.13^{+0.11}_{-0.10}$ & $380^{+9}_{-24}$ & $30^{+110}_{-110}$ & $0.000^{+0.005}_{-0.002}$ & $75^{+3}_{-3}$ \\
		&  &  &  &  &  & 5 & 764 & $1.2395657^{+0.0000012}_{-0.0000012}$ & $1.12^{+0.02}_{-0.02}$ & $380^{+9}_{-19}$ & $50^{+100}_{-140}$ & $0.0001^{+0.0007}_{-0.0006}$ & $113^{+5}_{-4}$ \\
		&  &  &  &  & 120 & 10 & 553 & $1.239566^{+0.000006}_{-0.000005}$ & $1.13^{+0.06}_{-0.08}$ & $380^{+10}_{-20}$ & $40^{+100}_{-120}$ & $0.000^{+0.002}_{-0.002}$ & $112^{+5}_{-5}$ \\
		&  &  &  &  &  & 20 & 446 & $1.23956^{+0.00003}_{-0.00001}$ & $1.12^{+0.15}_{-0.10}$ & $380^{+10}_{-30}$ & $30^{+120}_{-130}$ & $0.000^{+0.008}_{-0.002}$ & $112^{+6}_{-6}$ \\
		\hline 		&  &  &  &  &  & 5 & 2407 & $1.3153038^{+0.0000005}_{-0.0000006}$ & $1.289^{+0.005}_{-0.005}$ & $444^{+9}_{-9}$ & $100^{+70}_{-40}$ & $0.0000^{+0.0003}_{-0.0005}$ & $38.5^{+0.9}_{-0.8}$ \\
		&  &  &  &  & 40 & 10 & 1742 & $1.315303^{+0.000003}_{-0.000003}$ & $1.28^{+0.02}_{-0.02}$ & $449^{+10}_{-15}$ & $120^{+70}_{-70}$ & $-0.0004^{+0.0011}_{-0.0010}$ & $38.2^{+1.1}_{-0.9}$ \\
		&  &  &  &  &  & 20 & 1405 & $1.315296^{+0.000017}_{-0.000009}$ & $1.26^{+0.07}_{-0.04}$ & $452^{+9}_{-19}$ & $120^{+70}_{-60}$ & $-0.002^{+0.004}_{-0.002}$ & $38.0^{+1.1}_{-1.0}$ \\
		&  &  &  &  &  & 5 & 1204 & $1.3153039^{+0.0000010}_{-0.0000010}$ & $1.290^{+0.009}_{-0.009}$ & $444^{+13}_{-15}$ & $100^{+80}_{-70}$ & $-0.0001^{+0.0005}_{-0.0006}$ & $76^{+3}_{-2}$ \\
		1.3153039 & 1.289 & 442 & 75 & 0 & 80 & 10 & 871 & $1.315303^{+0.000005}_{-0.000005}$ & $1.29^{+0.03}_{-0.03}$ & $450^{+10}_{-20}$ & $13^{+70}_{-100}$ & $-0.000^{+0.002}_{-0.002}$ & $75^{+3}_{-3}$ \\
		&  &  &  &  &  & 20 & 702 & $1.315300^{+0.000024}_{-0.000019}$ & $1.27^{+0.10}_{-0.08}$ & $450^{+10}_{-40}$ & $110^{+80}_{-100}$ & $-0.001^{+0.006}_{-0.005}$ & $75^{+4}_{-3}$ \\
		&  &  &  &  &  & 5 & 802 & $1.3153038^{+0.0000016}_{-0.0000014}$ & $1.29^{+0.01}_{-0.01}$ & $440^{+10}_{-20}$ & $120^{+70}_{-100}$ & $-0.0000^{+0.0008}_{-0.0007}$ & $114^{+5}_{-4}$ \\
		&  &  &  &  & 120 & 10 & 581 & $1.315306^{+0.000006}_{-0.000008}$ & $1.30^{+0.04}_{-0.05}$ & $440^{+20}_{-30}$ & $90^{+90}_{-110}$ & $0.001^{+0.002}_{-0.003}$ & $112^{+1}_{-4}$ \\
		&  &  &  &  &  & 20 & 468 & $1.31530^{+0.00003}_{-0.00002}$ & $1.27^{+0.11}_{-0.10}$ & $445^{+15}_{-30}$ & $110^{+80}_{-110}$ & $-0.001^{+0.008}_{-0.006}$ & $112^{+6}_{-5}$ \\
		\hline 		&  &  &  &  &  & 5 & 2268 & $1.2236917^{+0.0000005}_{-0.0000005}$ & $1.286^{+0.005}_{-0.005}$ & $530^{+9}_{-11}$ & $150^{+70}_{-60}$ & $-0.0000^{+0.0004}_{-0.0006}$ & $38.6^{+1.0}_{-0.9}$ \\
		&  &  &  &  & 40 & 10 & 1642 & $1.223691^{+0.000003}_{-0.000003}$ & $1.28^{+0.02}_{-0.02}$ & $535^{+7}_{-15}$ & $190^{+40}_{-90}$ & $-0.0005^{+0.0012}_{-0.0011}$ & $38.2^{+1.2}_{-0.9}$ \\
		&  &  &  &  &  & 20 & 1325 & $1.223690^{+0.000008}_{-0.000008}$ & $1.28^{+0.04}_{-0.04}$ & $533^{+9}_{-17}$ & $160^{+60}_{-80}$ & $-0.001^{+0.003}_{-0.002}$ & $38.1^{+1.3}_{-1.2}$ \\
		&  &  &  &  &  & 5 & 1134 & $1.2236918^{+0.0000009}_{-0.0000008}$ & $1.287^{+0.009}_{-0.008}$ & $529^{+11}_{-17}$ & $160^{+60}_{-100}$ & $-0.0000^{+0.0006}_{-0.0005}$ & $76^{+2}_{-2}$ \\
		1.2236919 & 1.288 & 526 & 110 & 0 & 80 & 10 & 821 & $1.223692^{+0.000004}_{-0.000004}$ & $1.29^{+0.03}_{-0.03}$ & $529^{+13}_{-21}$ & $170^{+60}_{-100}$ & $-0.000^{+0.002}_{-0.002}$ & $75^{+3}_{-3}$ \\
		&  &  &  &  &  & 20 & 662 & $1.22369^{+0.00002}_{-0.00001}$ & $1.30^{+0.09}_{-0.07}$ & $522^{+18}_{-38}$ & $130^{+90}_{-100}$ & $0.001^{+0.006}_{-0.004}$ & $75^{+3}_{-3}$ \\
		&  &  &  &  &  & 5 & 756 & $1.2236919^{+0.0000014}_{-0.0000012}$ & $1.289^{+0.014}_{-0.012}$ & $525^{+14}_{-26}$ & $160^{+70}_{-130}$ & $0.0001^{+0.0007}_{-0.0006}$ & $113^{+5}_{-4}$ \\
		&  &  &  &  & 120 & 10 & 547 & $1.223694^{+0.000006}_{-0.000007}$ & $1.30^{+0.04}_{-0.05}$ & $520^{+20}_{-33}$ & $140^{+90}_{-120}$ & $0.001^{+0.003}_{-0.003}$ & $112^{+6}_{-5}$ \\
		&  &  &  &  &  & 20 & 442 & $1.22369^{+0.00003}_{-0.00002}$ & $1.29^{+0.12}_{-0.11}$ & $523^{+18}_{-48}$ & $140^{+80}_{-140}$ & $0.000^{+0.009}_{-0.007}$ & $112^{+5}_{-6}$ \\
		\hline 		&  &  &  &  &  & 5 & 2089 & $1.1076460^{+0.0000004}_{-0.0000003}$ & $1.125^{+0.008}_{-0.009}$ & $570^{+6}_{-10}$ & $90^{+140}_{-110}$ & $0.0000^{+0.0003}_{-0.0003}$ & $38.4^{+0.8}_{-0.8}$ \\
		&  &  &  &  & 40 & 10 & 1514 & $1.107646^{+0.000002}_{-0.000002}$ & $1.13^{+0.03}_{-0.03}$ & $569^{+6}_{-16}$ & $60^{+100}_{-100}$ & $0.0002^{+0.0008}_{-0.0008}$ & $38.2^{+1.4}_{-1.0}$ \\
		&  &  &  &  &  & 20 & 1222 & $1.107646^{+0.000008}_{-0.000006}$ & $1.13^{+0.07}_{-0.06}$ & $567^{+7}_{-17}$ & $50^{+120}_{-90}$ & $0.000^{+0.003}_{-0.002}$ & $38.0^{+1.3}_{-1.1}$ \\
		&  &  &  &  &  & 5 & 1045 & $1.1076460^{+0.0000007}_{-0.0000007}$ & $1.126^{+0.014}_{-0.014}$ & $567^{+8}_{-12}$ & $140^{+90}_{-160}$ & $0.0001^{+0.0005}_{-0.0005}$ & $76^{+2}_{-3}$ \\
		1.1076460 & 1.125 & 570 & 50 & 0 & 80 & 10 & 757 & $1.107647^{+0.000003}_{-0.000002}$ & $1.13^{+0.04}_{-0.03}$ & $568^{+9}_{-29}$ & $80^{+160}_{-260}$ & $0.0003^{+0.0016}_{-0.0010}$ & $76^{+3}_{-3}$ \\
		&  &  &  &  &  & 20 & 611 & $1.107646^{+0.000017}_{-0.000008}$ & $1.14^{+0.12}_{-0.10}$ & $565^{+10}_{-32}$ & $70^{+150}_{-150}$ & $0.000^{+0.006}_{-0.003}$ & $75^{+3}_{-3}$ \\
		&  &  &  &  &  & 5 & 696 & $1.1076461^{+0.0000011}_{-0.0000009}$ & $1.13^{+0.02}_{-0.02}$ & $564^{+11}_{-28}$ & $90^{+130}_{-220}$ & $0.0001^{+0.0006}_{-0.0006}$ & $113^{+5}_{-5}$ \\
		&  &  &  &  & 120 & 10 & 505 & $1.107646^{+0.000005}_{-0.000005}$ & $1.13^{+0.06}_{-0.09}$ & $565^{+11}_{-25}$ & $90^{+140}_{-220}$ & $0.000^{+0.002}_{-0.002}$ & $113^{+5}_{-5}$ \\
		&  &  &  &  &  & 20 & 407 & $1.107643^{+0.000019}_{-0.000007}$ & $1.11^{+0.14}_{-0.09}$ & $560^{+14}_{-39}$ & $90^{+190}_{-170}$ & $-0.001^{+0.007}_{-0.002}$ & $112^{+6}_{-6}$ \\
		\hline
		\hline
\label{tab:IM2}
\end{longtable}

\end{document}